\documentclass[svgnames,a4paper,twoside,12pt]{book}
\usepackage{tikz}
\usepackage{kpfonts}
\usepackage[utf8]{inputenc}
\usepackage[explicit]{titlesec}
\usepackage{array}
\usepackage{fancyhdr}
\usepackage{fancybox}
\usepackage[english]{babel}
\usepackage{latexsym} 
\usepackage{amsfonts}
\usepackage{amsmath}
\usepackage{amssymb}
\usepackage{longtable}
\usepackage{float}
\usepackage[bookmarks, colorlinks=false, pdftitle={Smart Resource Allocation}, pdfauthor={Pol Henarejos}, pdfsubject={Smart Resource Allocation}, pdfkeywords={PDF, LaTeX, hyperlinks, hyperref}]{hyperref}
\usepackage{lettrine}
\usepackage{caption}
\usepackage{algorithm}
\usepackage{algorithmicx}
\usepackage{algpseudocode}
\usepackage[a4paper]{geometry}

\fancypagestyle{plain}{ %
\fancyhf{} 
}
\newcommand*\chapterlabel{}
\titleformat{\chapter}
  {\gdef\chapterlabel{}
   \normalfont\sffamily\Huge\bfseries\scshape}
  {\gdef\chapterlabel{\thechapter\ }}{0pt}
  {\begin{tikzpicture}[remember picture,overlay]
    \node[yshift=-3cm] at (current page.north west)
      {\begin{tikzpicture}[remember picture, overlay]
        \draw[fill=LightSkyBlue] (0,0) rectangle
          (\paperwidth,3cm);
        \node[anchor=east,xshift=.9\paperwidth,rectangle,
              rounded corners=20pt,inner sep=11pt,
              fill=MidnightBlue]
              {\color{white}\chapterlabel#1};
       \end{tikzpicture}
      };
   \end{tikzpicture}
  }
\emergencystretch=\maxdimen
\titlespacing*{\chapter}{0pt}{50pt}{30pt}

\newenvironment{definition}[1][Definition]{\begin{trivlist}
\item[\hskip \labelsep {\bfseries #1}]}{\end{trivlist}}
\newcommand{\EX}[1] {{\mathbb{E}}\left\{{#1}\right\}}
\newcommand{\mub}{\boldsymbol{\mu}}
\newcommand{\phib}{\boldsymbol{\phi}}
\newcommand{\OO}[1]{\mathcal{O}\left(#1\right)}
\DeclareMathSymbol{\Pii}{\mathalpha}{letters}{"05}
\newcommand{\vc}[1]{{\bf #1}}

\newcommand{\PP}{\mathcal{P}}
\newcommand{\RR}{\mathcal{R}}

\begin{document}
\renewcommand{\thepage}{\roman{page}} 
\setcounter{page}{1}
\begin{titlepage}

\newcommand{\HRule}{\rule{\linewidth}{0.5mm}}
\center
\textsc{\LARGE Universitat Politècnica de Catalunya}\\[1.5cm] 
\textsc{\Large European Master of Research on Information and Communication Technologies}\\[1cm]
\textsc{\large Centre Tecnològic de Telecomunicacions de Catalunya}\\[1.5cm]
\HRule \\[0.7cm]
{ \huge \bfseries Smart Resource Allocation}\\[0.2cm]
{ \Large Beyond the optimum }\\[0.2cm]
\HRule \\[1.5cm]
\begin{minipage}{0.4\textwidth}
\begin{flushleft} \large
\emph{Author:}\\
Pol \textsc{Henarejos} 
\end{flushleft}
\end{minipage}
~
\begin{minipage}{0.4\textwidth}
\begin{flushright} \large
\emph{Supervisor:} \\
Dr. Ana \textsc{Pérez Neira} 
\end{flushright}
\end{minipage}\\[4cm]
\begin{tabular}{m{0cm}m{3cm}m{3cm}m{3cm}}
\, &
\includegraphics[width=0.15\textwidth]{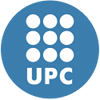} &
\, &
\includegraphics[width=0.15\textwidth]{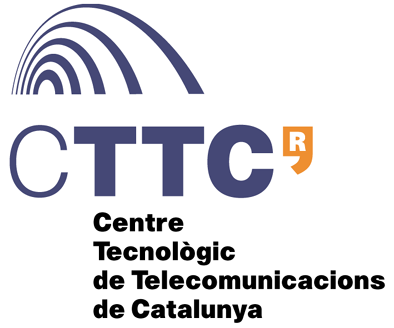} 
\\
\end{tabular}\\[1cm]
{\large \today}\\[3cm]
\end{titlepage}
\clearpage
\null
\vfill
\begin{center}
Pol Henarejos Hernández, © 2012\\
All Rights Reserved.
\end{center}
\newpage
\thispagestyle{empty}
   
\newpage

\thispagestyle{empty}
~\\[15cm]
\makebox[16cm][r]{\lq\lq \textit{The value of an idea lies in the using of it.}\rq\rq}
\makebox[16cm][r]{Thomas A. Edison}

\newpage
\thispagestyle{empty}
\newpage
\chapter*{Acknowledgements}
\addcontentsline{toc}{chapter}{Acknowledgements}
I would like to thanks with this work to my wife, Sara. She is my support when I fall and gives me the reasons to continue and never give up. I love you.

Second, I also would thank to my family. They are always to my side. They interest for all my jobs and specially with this master thesis. Thanks, thanks, thanks.

Third, to my CTTC colleges. Most of them are at the same situation as me. And they also know the hard task of combining the normal job and the PhD. pursue. There are not enough coffees to thank you.

Finally, and not less important, to my advisor. She guided me for many years and advised me to get my results, even before ending my degree. Thank you, Ana.

And if you are reading this and you do not know who am I, also thank you. Probably you arrived here motivated by finding solutions to your problems. I bet that I have not any of them. But I hope to wake up an interest on you about the passionate world of implementing theoretical technologies to real devices.

\chapter*{Abstract}
\addcontentsline{toc}{chapter}{Abstract}

\noindent{\bf Title:} Smart Resource Allocation: Beyond the Optimum\\*
\noindent{\bf Author:} Pol Henarejos Hernández\\*
\noindent{\bf Affiliation:} Centre Tecnològic de Telecomunicacions de Catalunya (CTTC)
\noindent{\bf Advisor:} Ana Pérez Neira

\noindent{\bf Abstract:}
Dealing with broadcast scenarios has become a relevant topic in the scientific community. Because of interference, resource management presents a challenge, specially when spatial diversity is introduced. Many researches presented theoretical benchmarks, simplifications and low complex schemes, but in fact it is difficult the real implementation. 

The major part of current works propose iterative solutions, which are far away of feasible results. Since the problem is not convex, iterative solutions are prohibitive. Moreover, they always require the full knowledge of the channel state information. Hence, the feedback channel is often unaffordable and makes impossible to carry a huge amount of information.

The present work aims to fill this gap presenting a novel scheme, from the theoretical framework to realistic scheme. It introduces the solution of the maximization of the sum rate in the broadcast scenario with multiple antennas at the transmitter. This solution aims to be realistic and to distribute the complexity between the base station and user equipments.

Also, due to its construction, it opens the door to be compatible with LTE standards with no relevant changes. Thus, it allows to combine the resource allocation with scheduling tasks in a LTE environment, fulfilling the feedback requirements and maximizing the sum rate of the system.

{\bf Keywords:} resource allocation, MIMO, OFDMA, broadcast, LTE.

\chapter*{Notation}
\addcontentsline{toc}{chapter}{Notation}

\begin{longtable}{rl}
$a$ & Scalar, constant, parameter.\\[0.5cm]
$\vc{a}$ & Vector, in column way.\\[0.5cm]
$\vc{A}$ & Matrix.\\[0.5cm]
$\vc{a}^T$ & Transposed vector $\vc{a}$. It is equal to a row vector.\\[0.5cm]
$\vc{a}^H$ & Hermitian vector.\\[0.5cm]
$\vc{A}^T$ & Transposed matrix.\\[0.5cm]
$\vc{A}^*$ & Conjugate matrix.\\[0.5cm]
$\vc{A}^H$ & Hermitian matrix.\\[0.5cm]
$|a|$ & Magnitude of $a$.\\[0.5cm]
$\|\vc{a}\|$ & Euclidean norm of $\vc{a}$.\\[0.5cm]
$\EX{\mathcal{X}}$ & Expectation of random variable $\mathcal{X}$. It is equal to\\& $\int\limits_{-\infty}^{+\infty}xf_{\mathcal{X}}(x)\,dx$.\\[0.5cm]
$\mathbb{R}$ & Real numbers group.\\[0.5cm]
$\mathbb{R}^+$ & Real positive numbers group.\\[0.5cm]
$\mathbb{R}^n$ & Group of $n$ real components vectors.\\[0.5cm]
$\mathbb{R}^{n\times m}$ & Group of $n$ by $m$ real matrices.\\[0.5cm]
$\mathbb{C}$ & Complex numbers group.\\[0.5cm]
$\mathbb{C}^n$ & Group of $n$ complex components vectors.\\[0.5cm]
$\mathbb{C}^{n\times m}$ & Group of $n$ by $m$ complex matrices.\\[0.5cm]
$\arg\max_x f(x)$ & Value of $x$ that maximizes the $f(x)$ function.\\[0.5cm]
\end{longtable}

\chapter*{Acronyms}
\addcontentsline{toc}{chapter}{Acronyms}

\begin{longtable}{rl}
\textsf{3G} & \textit{Third Generation}.\\[0.5cm]
\textsf{3GPP} & \textit{Third Generation Partnership}.\\[0.5cm]
\textsf{AWGN} & \textit{Additive White Gaussian Noise}.\\[0.5cm]
\textsf{BC} & \textit{Broadcast}.\\[0.5cm]
\textsf{BS} & \textit{Base Station}.\\[0.5cm]
\textsf{CDMA} & \textit{Code Division Multiple Access}.\\[0.5cm]
\textsf{CQI} & \textit{Channel Quality Indicator}.\\[0.5cm]
\textsf{CSI} & \textit{Channel State Information}.\\[0.5cm]
\textsf{CSIT} & \textit{CSI at the Transmitter}.\\[0.5cm]
\textsf{DPC} & \textit{Dirty Paper Coding}.\\[0.5cm]
\textsf{EDGE} & \textit{Enhanced Data rates for GSM Evolution}.\\[0.5cm]
\textsf{DSL} & \textit{Digital Subscriber Line}.\\[0.5cm]
\textsf{GPRS} & \textit{General Packet Radio System}.\\[0.5cm]
\textsf{GSM} & \textit{Global System for Mobile communications}.\\[0.5cm]
\textsf{IEEE} & \textit{Institute of Electrical and Electronics Engineers}.\\[0.5cm]
\textsf{KKT} & \textit{Karush-Kuhn-Tucker}.\\[0.5cm]
\textsf{LTE} & \textit{Long Term Evolution}.\\[0.5cm]
\textsf{MAC} & \textit{Multiple Access}.\\[0.5cm]
\textsf{MIMO} & \textit{Multiple Input Multiple Output}.\\[0.5cm]
\textsf{MISO} & \textit{Multiple Input Single Output}.\\[0.5cm]
\textsf{MMSE} & \textit{Minimum Mean Square Error}.\\[0.5cm]
\textsf{MOB} & \textit{Multiuser Opportunistic Beamforming}.\\[0.5cm]
\textsf{OFDM} & \textit{Orthogonal Frequency Division Multiplexing}.\\[0.5cm]
\textsf{OFDMA} & \textit{Orthogonal Frequency Division Multiple Access}.\\[0.5cm]
\textsf{P2P} & \textit{Point to Point}.\\[0.5cm]
\textsf{PRB} & \textit{Physical Resource Block}.\\[0.5cm]
\textsf{QoS} & \textit{Quality of Service}.\\[0.5cm]
\textsf{SIMO} & \textit{Single Input Multiple Output}.\\[0.5cm]
\textsf{SISO} & \textit{Single Input Single Output}.\\[0.5cm]
\textsf{SNIR} & \textit{Signal Interference Noise Ratio}.\\[0.5cm]
\textsf{SNR} & \textit{Signal to Noise power Ratio}.\\[0.5cm]
\textsf{UE} & \textit{User Equipment}.\\[0.5cm]
\textsf{UPA} & \textit{Uniform Power Allocation}.\\[0.5cm]
\textsf{ZF} & \textit{Zero Forcer}.\\[0.5cm]
\end{longtable}
\tableofcontents
\listoftables
\addcontentsline{toc}{chapter}{List of tables}
\listoffigures
\addcontentsline{toc}{chapter}{List of figures}
\newgeometry{marginparwidth=120pt,includemp}
\chapter{Introduction}
\renewcommand{\thepage}{\arabic{page}} 
\setcounter{page}{1}
\makebox[16cm][r]{\lq\lq \textit{Information is the resolution of uncertainty. }\rq\rq}
\makebox[16cm][r]{Claude E. Shanon}\\[1cm]
\lettrine[lines=3]{T}{his} work is a deep study on resource management in broadcasting scenarios. Concretely, the focus is on multiuser environments, what are the challenges and how service providers can manage them. 

\section{It is revolution}

Nowadays our society is suffering one of the major revolutions in his history. It is qualified as information revolution. People can access to the information around the globe instantly. In almost 15 years the society has passed from the use of desk phones to smartphones; from the use of personal computers as work tools to tablets for consuming a huge amount of information. In these few years, many challenges have appeared because of population demands. At the beginning only few users used GSM mobiles. This technology was conceived to carry voice in real time and short text messages.\marginpar{
\includegraphics[width=1.1\marginparwidth]{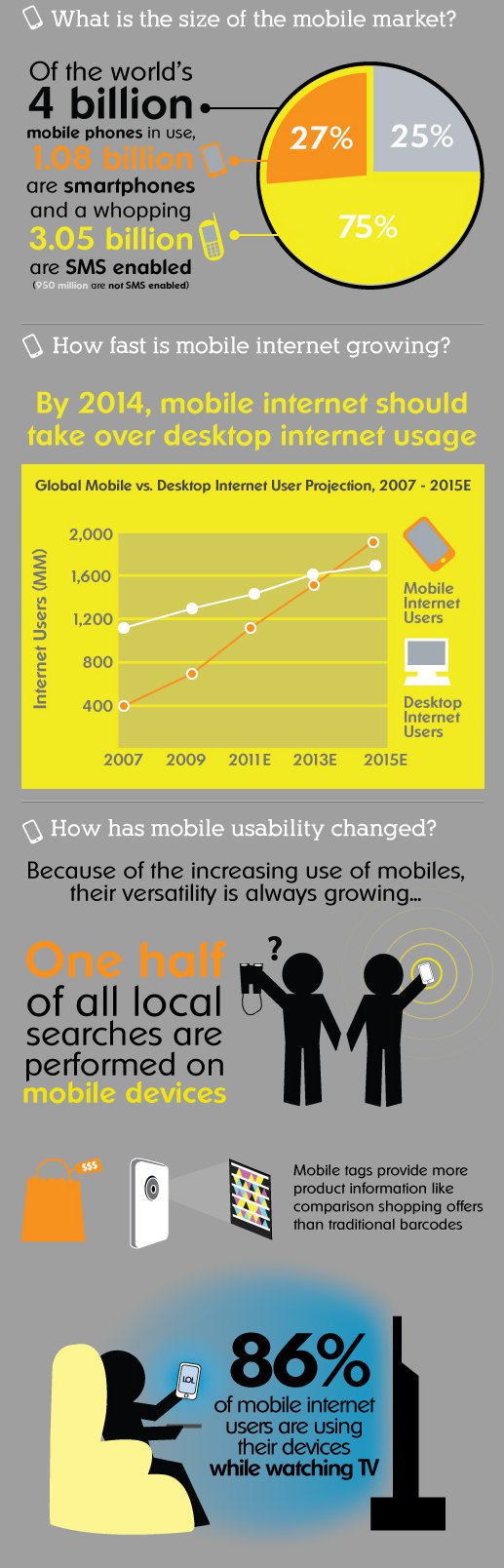}
\captionof{figure}{\footnotesize Usage of mobile devices}
} There were some failed protocols to carry data, such as GPRS or EDGE, over voice circuits but the extended use of data circuits came with the 3G technologies and the irruption of the smartphones into the market.

Each technology offered new services which most of them were adopted, demanding at the same time more services and opening more challenges at the horizon. This is one of the keys that made possible the revolution. Nowadays our society demands more mobility and user experiences. It is the society of sharing, of accessing the content whenever and of carrying its personal data with itself.

The population has more and more devices connected to the Internet and Internet is now more accessible. IPv6 born in order to increase the directionality of the net. When Internet was designed, no one thought on the potential and 4 millions of directions were enough. But nowadays they are not. In the GSM years, a bitrate of few kilobits were enough but today they are not. Services, subscribers, devices... all of them have grown up exponentially and the dimension of the telecommunications needs to be redefined.

\marginpar{
\includegraphics[width=1\marginparwidth]{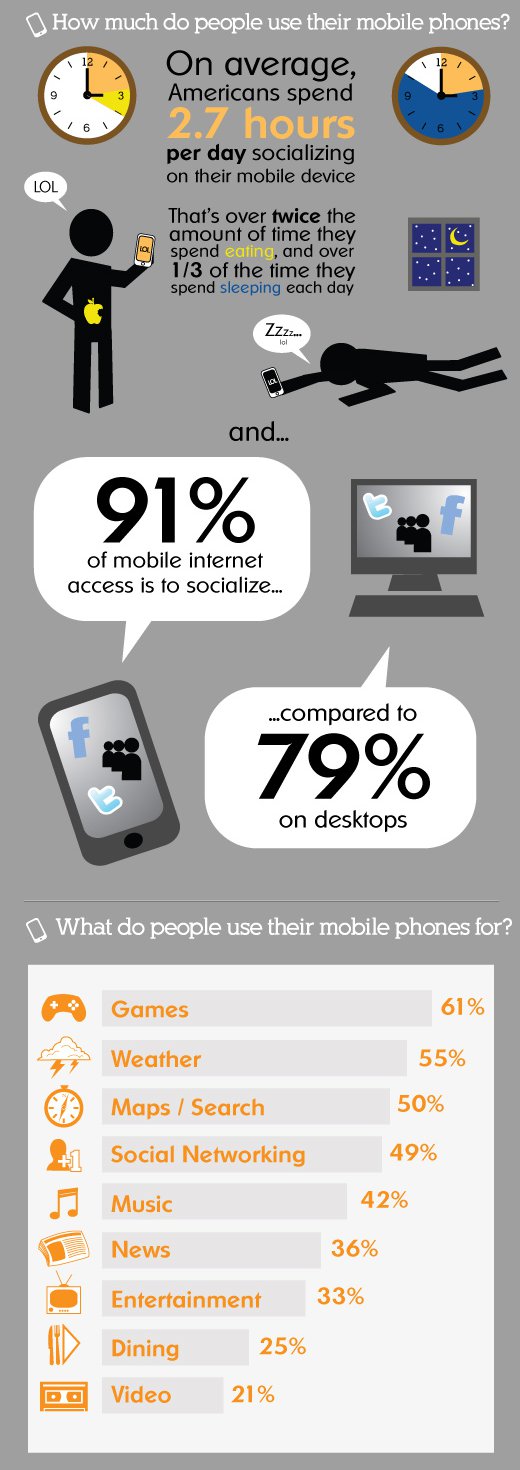}
\captionof{figure}{\footnotesize Usage of mobile devices}
} 

However, physical resources did not grow up at the same speed of demands and every day is more imperative to manage them in a smart way. Spectrum licenses and bandwidths are so restrictive that it is necessary to use other kind of resources. Spatial diversity has been presented as an effective way to deal with the increasing demand. Despite of it is an expensive technology, it allows to increase the bitrate, improve the bit error rate or serve more users with no \restoregeometry extra resources. More equipments are manufactured with more than one antenna and this technology is more adopted by providers, operators and subscribers. Standards such as IEEE 802.11n or LTE include multiantenna techniques as their main points and they response to the needs of the market.

Resource management needs to be effective not only with these physical resources but the traffic loads. To deal with many users whose interferences difficult the reliable communication, there are some lines of research that propose different philosophies. On one hand, there are those which reduce the interference to the minimum. On the other, there are those that allow interference and choose which are the less interfered. The major part of these works propose solutions based on iterative and expensive searches, which are impossible to put into the practice.

It introduces an ambitious challenge. Avoiding exhaustive searches is a major constraint if it is desirable its implementation and beside spatial diversity and multiuser environments the problem becomes unaffordable.

\section{Structure}
This work is structured as follows:
\begin{itemize}
\item Chapter 2 presents the current state of the art. It introduces some of the more extended currents. In this chapter many concepts are also described and analysed. In some of them lie the clues to understand the problem and why it is so hard to find the optimal solution.
\item Chapter 3 introduces the proposed scheme. It departs from the more general approach and reduces and simplifies it to be affordable. It describes the theoretical framework and also the algorithms that are implementable.
\item Chapter 4 shows graphical results and discusses some relevant aspects and how they determine the results.
\item Chapter 5 finally presents the conclusions and describes which could be the way to be followed.
\end{itemize}

\section{Derived works}

For the research of this work the following publications were been accepted:

\begin{itemize}
\item Journal paper

The following journal is under the second round of revision:
\begin{itemize}
\item P. Henarejos, A. Pérez Neira, V. Tralli, and M. A. Lagunas, \emph{Low-Complexity Resource Allocation with Rate Balancing for the MISO-OFDMA Broadcast Channel}, submitted to Elsevier Signal Processing.
\end{itemize}
\item International conferences
\begin{itemize}
\item V. Tralli, P. Henarejos, A. Pérez-Neira, \emph{A low complexity scheduler for multiuser MIMO-OFDMA systems with heterogeneous traffic}, in proceedings of International Conference on Information Networking (ICOIN 2011), 26-28 January 2011, Kuala Lumpur (Malasya).
\item V. Tralli, A. Pérez-Neira, P. Henarejos, \emph{A cross-layer scheduling strategy for the downlink of a MIMO-OFDMA system with heterogeneous traffic}, in Proceedings of the 9th MCM (COST 2100), 28-30 September 2009, Vienna (Austria).
\item A. Pérez-Neira, P. Henarejos, V. Tralli, M. Lagunas, \emph{A low complexity Space-Frequency Multiuser Scheduling Algorithm}, in Proceedings of NEWCOM++, ACoRN Joint Workshop, 30-1 April 2009, Barcelona (Spain).
\item A. Pérez-Neira, P. Henarejos, V. Tralli, M. Lagunas, \emph{A low complexity space-frequency multiuser resource allocation algorithm}, in Proceedings of International ITG Workshop on Smart Antennas (WSA 2009), 16-19 February 2009 Berlin, (Germany).
\end{itemize}
\end{itemize}

\chapter{Where we are. Where we go.}
\makebox[16cm][r]{\lq\lq \textit{I have the capacity of being more wicked than any example that man could set me.}\rq\rq}
\makebox[16cm][r]{James C. Maxwell}\\[1cm]
\lettrine[lines=3]{M}{any} lines of research focus on different aspects of the problem of resource management in broadcast scenarios\footnote{Where there are one transmitter and multiple receivers.}. As it will be shown later, the problem lies into maximization of the sum rate. This work focuses specially on the control of the power to conform the scenario and establishes different channel capacities. Nevertheless, it would not be a problem in single user P2P environments, where more power means more capacity. In multiuser scenarios, giving more power to one user means that others are receiving more interference and hence their capacity is reduced. The problem is how to balance this power to obtain the maximum sum rate. Once it is achieved, each user's rate is adjusted to this capacity and the information is transmitted using this power.

It is further clear that the channel capacity depends on the bandwidth, on the power and on the precoder. Therefore there are multiple physical variables to be optimized in order to obtain the solution under the scenario requirements. Additionally the assignment of resources to users has to be done. Moreover, since there are usually more users than available resources, there is another previous step in this process: before assigning the resources to users, it is necessary to select a reduced group of users which will compete in the resource assignment. And again, depending on the number of users in this set the interference will be different and will affect to sum rate. After all, the resulting optimizing problem is non convex.

Summarizing all these aspects, the whole resource management involves the following steps:
\begin{enumerate}
\item Select a group of users.
\item Assign physical resources to users of the group.
\item Adjust power for each assigned resource.
\item Compute the objective function and repeat these steps.
\end{enumerate}

As the reader can see, there exist multiple permutations and combinations which make the problem unfeasible, even for a low number of users. In this way, Costa in \cite{Costa1983} developed the DPC to control the interference and achieve sum capacity. Users are ordered and decoded successively to achieve sum capacity. However it requires a previous user group selection and its complexity is prohibitive. 

In recent years, researchers focus on reducing the complexity but maintaining the benefits. However, even when complexity can be reduced using simplifications, in most cases an exhaustive search is imperative.

Jointly with \cite{Costa1983}, authors of \cite{Cai01,Bre09,Tse01,Gol01} study different aspects of broadcast scenarios. In the case of \cite{Cai01}, the authors analyse several solutions in the theoretical frame to obtain the optimal solution. David Tse in \cite{Tse01} propounds the paradigm of MIMO scenarios. In these cases, Tse studies the behaviour of the vector inputs and outputs in presence of AWGN and the reciprocity of the analogue scenarios. Finally, \cite{Gol01} establishes a closed equation which relates MAC and broadcast scenarios.

Because of duality proved in \cite{Tse01,Gol01}, some authors solve the problem first in the MAC scenario and then convert the results to the equivalent broadcast scenario. Many other works as \cite{Cio02,Han05,Gian01,WYu01,Gian06,Yu06}, present the problem and solve it applying Lagrangian duality. In these cases, a dual and convex problem is proposed, which satisfies all the constraints of the original problem. The solution of the dual problem gives an upper bound, as well as reduces the complexity because of its convexity and, in some cases, it corresponds to the optimal value.

\section{Reaching sum capacity}

Reaching sum capacity is therefore one of the main goals that scientific community proposes. As mentioned before, the complexity is one of the drawbacks. Additionally, OFDMA becomes a more generalized scheme to offer communications in multiuser scenarios. Since each subcarrier can be treated as a single narrowband channel, the solutions explained before can be applied in each subcarrier. Hence, the complexity is even higher, as the procedure is repeated for all subcarriers. This additional degree of freedom motivated the authors of \cite{Noss01,Ott08,Wei08,Lat07,Kou06,Wan07,Tej01,wsa09,LHo09} to find different solutions. For instance, in \cite{Tej01} the structure of the channel matrix is diagonal and exploited to reduce the complexity and make more practicable the implementation. In this case, the solution of spatial and frequency diversities is jointly found. In \cite{Ott08,Wei08,Kou06,Wan07}, suboptimal techniques are proposed in order to reduce complexity.

As mentioned before, the power assignment also determines the channel capacity and takes a relevant role. Choosing the user set implies to known a priori which amount of power will be granted to each user, and hence the quantity of interference. But before computing the amount of power to be distributed it is necessary to give the user set. This is a mutual coupled problem that is still open. As there is no solution yet, some authors prefer to keep the power constant\footnote{Also known as UPA.} and dedicate efforts to choose the optimal user set. In \cite{Wan07}, the authors focus on the user and frequency assignment problem, whereas power is uniformly allocated.

\section{More than one antenna}
Schemes with multiple antennas certainly improve the overall throughput. But having more than one antenna means that there is an implicit interference. Even though there is only one user in the system, the information propagated through the antennas suffers interference if it is not properly manipulated. The optimal solution is adding a constraint where the interference is null. This is equivalent, in terms of mathematics, to diagonalize the channel matrix. In these cases, the information of each antenna is propagated in a way that does not interfere to the rest. This reasoning can be applied to multiuser schemes, as \cite{Moon2010,Joung2007} do.

Nevertheless, in multiuser environments there are still interference among all users. Transmitting multiple information to several users can be harmful as some users can receive tainted payload. This can be easily avoided placing a precoder before transmitting. Hence, the information can be packeted spatially and sent with no interference. In this sense, publications such as \cite{Spencer2004} play with the ZF concept to obtain a signal with no interference. Other strategies relax the constraints and allow some interference. This is the case of MMSE, which departs from the premise that some interference may be profitable because does not affect to the capacity of others and which is based on Costa's DPC.

The other concept in the precoding design is called Opporunistic Beamforming, where the randomness of channel conditions can be exploited to perform the user selection. In other words, in a particular time slot some users will have better channel conditions and not others. In another time slot, could be plausible that those users may have worse channel conditions but not the others. Thus, this scheme selects in each time which are the better users to incorporate to the scheduler. Viswanath et. al show in \cite{Tse02} how this philosophy changes the previous concepts, where channel must be defeated. In \cite{Zor01} the authors also analyse the impact of this criteria and propose that the more users the better the results are. Moreover, the scheme can achieve the sum capacity asymptotically for a large number of users.

\section{Considerations}

In the following section some important aspects for this work are introduced.

\subsection{Too expensive}
The major drawback of multiuser scenarios is that there is an implicit search. The efforts are focused specially on reducing the dimensionality of this search. This fact motivates iterative algorithms. As it is complex to perform a jointly selection of resources, some works decouple the problem into a set of two subproblems: first solving the unitary precoder matrix and then the power solution. 

In this sense, \cite{Tej01} and more references therein propose a recursive maximization. For each time slot, first the covariance of the precoder is solved and this result is used for the power allocation. After that, the covariance is solved again using this power allocation scheme and so on and so forth until the tolerance is reached. At this point, the optimal values are found but it is necessary a huge computational consumption. This iterative approach is in practice unfeasible and it is not possible to implement in real-time devices.

Nevertheless, as mentioned before, opportunistic beamforming schemes can reduce the complexity and it can be reflected specially during the beamforming allocation stage. For instance, in \cite{Kan08} authors explain how the complexity can be further reduced, passing from the initial one $O\left(M\left(Kt\right)^t\right)$ to the more reduced $O\left(MKt\right)$, where $M$ stands for the number of subcarriers, $K$, for the number of UEs and $t$, for the number of transmitter antennas. The authors describe which procedure is applied and specially which is the gap between the optimal and suboptimal results. They show that the complexity decreases but the sum rate is not so much penalized. Hence, the exhaustive search can be simplified but maintaining the overall rate.

\subsection{Relaxing constraints}

In despite of the hard task of reducing the computational complexity, there exist other approaches. The ergodicity is widely used in many algorithms since it relaxes the constraints and makes no necessary to iterate the maximization algorithm on each time slot. This approach is not new and has been used in other scopes, such as CDMA, as in \cite{Kaya2005} authors narrate. But authors of \cite{Won07,WonEvaBook} extend this reasoning to the OFDMA schemes. In \cite{WonEvaBook}, ergodic algorithms are presented in such a way of reducing complexity and feasible implementation. Even though they focus on SISO multicarrier scenarios, the present work introduces the spatial dimension using the same approach. However, as explained before, spatial scenarios introduce interference among users and it is not straightforward to extend it.

\subsection{Closing the loop}

Beside computational complexity there is the feedback complexity. Both concepts are closely related. A high degree of variables passed from the UEs to the BS implies more parameters to be computed and hence more variables to be optimized. Thus, another handicap appears: the degree of feedback. Most of mentioned works require the full knowledge of the channel state by the transmitter before conveying. This means that a lot of information shall be transmitted by the users to the BS. For example, a typical system of $64$ subcarriers, MIMO $2x2$, $1$ms of slot and $8$-bits for each codeword, requires more than $2$Mbps of continuous flow for each user, all of them corresponding only to the channel coefficients. This amount is unaffordable by any system and makes no possible its implementation. 

LTE is the first major standard to include some feedback parameters and it does in a very small amount and privileged users. In \cite{TS36213} authors show that there are three kind of parameters: the rank of the precoder, the codebook index of the precoder and the channel quality indicator. Depending on the transmission scheme used and the bandwidth, it requires around $8$Kbps\footnote{It depends on a multiple number of parameters but always is Kilobits per second of magnitude order.}. Hence, how can these small number of bits provide a reliable channel state information? Firstly, doing temporal and frequency simplifications. In time, the CSI is assumed constant during all subframe ($1$ms). In frequency, subcarriers are grouped in chunks of $12$ subcarriers (one \emph{PRB}), meaning that all of them have the same CQI value. Finally, it is important to remark that it is not strictly the CQI, but a kind of metric of it. Hence, the full CSI is compressed in a metric which provides the same useful information for the transmitter and therefore the feedback amount is further reduced.

\subsection{Privileges or equanimity}

Fairness is a recurrent and discussed topic in the literature. But, how can an abstract concept be defined? How can it be applied to information theory? Certainly, it is not an easy task and depends on the current epoch. In 1984, authors of \cite{Raj84} tried to define what is fairness and what is not. This interesting article describes intuitively which should be a fairness index of a system, even though at this time did not exist wireless commercial communications. This index, nowadays known as Jain's index, propounds the following statement:

\begin{definition}
A set $\left\{x_i\right\}$ of $n$ variables is fairly distributed if 
\begin{equation}
f(x)=\frac{\left|\sum\limits_nx_i\right|^2}{n\sum\limits_nx_i^2}=1.
\end{equation}
\end{definition}

It is clear that when all $\left\{x_i\right\}$ have the same value, this index is 1. If all values are $0$ except one, this index is asymptotically $0$.

This index analyses the distribution of the elements. When all elements are equally distributed the index is maximum and when they are all dispersed, the index is minimum. Hence, it gives a notion on how the system is fair or not. However, this metric evaluates the fairness in terms of equality. 

There are many other indices with other concepts of fairness. Previous to Jain's index, there were the following indices:
\begin{enumerate}
\item Variance index: $\sigma_x=\frac{1}{n-1}\sum_n\left(x_i-\mu\right)^2$, where mean $\mu=\frac{1}{n}\sum_nx_i$.
\item Coefficient of variation: $CoV=\frac{\sigma_x}{\mu_x}$.
\item Min-max ratio: $mmr=\frac{\min_i\left\{x_i\right\}}{\max_j\left\{x_j\right\}}=\min_{i,j}\left\{\frac{x_i}{x_j}\right\}$.
\end{enumerate}

These indices could present some advantages years ago, but nowadays are quite obsolete. The network has been increasing year by year and classes have appeared. It means that the resources are not distributed equally, but depending on users' demands. In DSL lines, there are several classes depending on the speed of the link. In 3G connections, classes are set on budgets of available data for downloading at maximum rate. Hence, it seems clear that the system is fair if all users get what they purchased.

In this sense, authors of \cite{Rodrigues2011} use a modified version of Jain's index. It is summarized as follows:

\begin{equation}
f(x)=\frac{\left|\sum\limits_n\phi\left(x_i\right)\right|^2}{n\sum\limits_n\phi\left(x_i\right)^2}.
\end{equation}
where $\phi\left(x_i\right)=\frac{x_i}{x_i^{req}}$ and $\left\{x_i^{req}\right\}$ is the magnitude requested. Again, if all users obtain what they requested, $\phi\left(x_i\right)$ coefficients are $1$ and the index is also $1$. And it is under $1$ if some user does not get what is requested.

This approach gives more flexibility in the constraints and accepts different classes and requirements. Certainly, it still accomplishes the postulates in \cite{Raj84} and, since it is derived from Jain's index, all assumptions and properties are valid. Notwithstanding, it is difficult to include Jain's index in maximization problems since its derivative contains quadratic terms and powers of four, which difficults finding the solution. Thus, fairness indices are introduced into the formulation of the problem in such a way that they appear as adaptive formulas, but not in the maximization problem. This is done typically adding some constraint that depends directly with a parameter that is computed adaptively and it is a function of fairness index. See \cite{Rodrigues2011} for a complete example.

In addition to this procedure, there are other publications that include fairness maximization into the problem. In the work \cite{Ismail2009} the fairness term appears in the maximization statement in a way of logarithm. Instead of maximizing the sum rate, it maximizes the sum of the logarithms of individual rates. That is equivalent to maximize the logarithm of the product sequence of individual rates. Hence, if the distribution is not uniform, the maximization is penalized. This behaviour is analogue to the original Jain's index but can be introduced into the main problem and solved using traditional utilities.

In contrast of fairness concept, it is possible to include class services or QoS as constraints of the maximization problem. This can be done simply adding constraints where individual rates shall be below or above certain amount. This is reflected in a deep analysis in \cite{WonEvaBook}. QoS often is not only a rate requirement but a complete profile which contains rate, bit error rate, delay and many other parameters. However, this approach can be a first attempt to include QoS specifications, typical from higher layer, into the physical layer. 

\section{Ambitions}

As mentioned before, resource allocation is a wide problem, which comprises precoder selection, power allocation, beam allocation, fairness stability, maintained complexity, feedback reduction and multiuser selection. These aspects are in fact an interesting challenge and many authors proposed a wide variety of solutions, more or less effective, but it is undeniable that is a glowing topic.

This work aims to fit these aspects in a proper way and give its solution. To summarize, the problem is presented as follows:

\begin{itemize}
\item Scenario with $K$ users, $t$ transmitter antennas and $M$ active subcarriers.
\item The objective is the maximization of the sum rate.
\item QoS and rate constraints.
\item Power constraint.
\end{itemize}

In the following chapter, the problem and the solutions are provided. Also, a description of the original problem is introduced and how the complexity can be reduced.

\chapter{Smart Resource Allocation}

\makebox[16cm][r]{\lq\lq \textit{I have had my results for a long time. }}
\makebox[16cm][r]{\textit{But I do not yet know how I am to arrive at them.}\rq\rq}
\makebox[16cm][r]{Carl F. Gauss}\\[1cm]
\lettrine[lines=3]{S}{ystems}, such as OFDM, with $M$ subcarriers are considered. The transmitter, which can be a BS, is equipped with $t$ antennas and delivers signals to $K$ receiving terminals or users. According to the MIMO-OFDM BC model, each receiver is equipped with $r_k$ antennas, $k=1,\ldots,K$.
The received signal ${\bf y}_{k,m} \in \mathbb{C}^{r_k\times 1}$ for user $k$ at the $m$-th subcarrier, $m=1,\ldots,M$,  is expressed as
\begin{equation}
{\bf y}_{k,m}={\bf H}_{k,m} {\bf x}_m + {\boldsymbol {\omega}}_{k,m}
\label{eq:my}
\end{equation}
where ${\bf x}_{m} \in \mathbb{C}^{t\times 1}$ is the transmitted signal, ${\bf H}_{k,m} \in \mathbb{C}^{r_k\times t}$ is the channel matrix and ${\boldsymbol {\omega}}_{k,m} \in \mathbb{C}^{r_k\times 1}$ is the zero-mean circularly symmetric complex Gaussian noise with covariance matrix $\sigma^2_\omega{\bf I}_{r_k}$,  which is assumed to be uncorrelated on the different subcarriers. Eq. (\ref{eq:my}) can be rewritten in a compact form as
\begin{equation}
\tilde{\bf y}_{k}=\tilde{\bf H}_{k} \tilde{\bf x} + \tilde{\boldsymbol {\omega}}_{k}
\end{equation}
where $\tilde{\bf y}_k= \left[ {\bf y}_{k,1}^T \cdots {\bf y}_{k,M}^T \right]^T$, $\tilde{\bf H}_k= {\rm diag}\left[ {\bf H}_{k,1}, \ldots, {\bf H}_{k,M} \right]$, $\tilde{\bf x}= \left[ {\bf x}_1^T \cdots {\bf x}_M^T \right]^T$ and $\tilde{\boldsymbol {\omega}}_k= \left[ {\boldsymbol {\omega} }_{k,1}^T \cdots {\boldsymbol {\omega}}_{k,M}^T \right]^T$.
The outlined MIMO-OFDM system can also be viewed as classical MIMO BC with channel matrix $\hat{\bf H}= \left[ \tilde{\bf H}_1^T \cdots \tilde{\bf H}_K^T \right]^T$. The vector of transmitted signal $\tilde{\bf x}$ is then given by the superposition of signals $\tilde{\bf s}_k= \left[ {\bf s}_{k,1}^T \cdots {\bf s}_{k,M}^T \right]^T$ intended for user $k$, $\tilde{\bf x}=\sum_{k}\tilde{\bf s}_{k}$.

In this scenario, the problem of building the signals $\tilde{\bf s}_k$ in order to maximize the sum-rate with a constraint on the ratios between the user-specific rate $R_k$ and sum-rate $\Gamma=\sum_kR_k$, for $k=1,\ldots,K$ is considered. This problem is known as rate balancing \cite{Tej01} and each constraint can be viewed as a QoS constraint on the rate requirements coming from the upper-layers of the system. Hence, the problem is formulated as
\begin{equation}
\begin{split}
\label{eq:ratebal}
& \max_{\Gamma} \Gamma\\
s.t. \;\; & R_k=\phi_k\Gamma , \;\; \forall k \\
& (R_1,\ldots,R_K) \in \mathcal{C} (\hat{\bf H},\overline{P})
\end{split}
\end{equation}
where $\overline{P}$ is the total power budget and $\mathcal{C} (\hat{\bf H},\overline{P})$ is the capacity region of the BC. The first constraint indicates that user $k$ has to obtain a fraction $\phi_k$ of the achievable sum-rate; thus, $\sum_k \phi_k =1$.
The second constraint indicates that the vector of rates must lay in the capacity region corresponding to channel $\hat{\bf H}$ and power budget $\overline{P}$. Sum capacity of the MIMO BC channel has not a closed form and can be expressed using the MAC-BC duality. Nevertheless, it can always be compacted using the form of $\log\left(1+X\right)$, where $X$ is the corresponding signal-to-noise ratio, in the case of SISO, SIMO or MISO schemes.

QoS constraints fix which amount sum rate shall be granted to each user. In this sense, fairness appears in this terms of classes. Depending on the $\phi_k$ parameter, it is possible to adjust several users classes to fulfil different rates. Thus, the fairness is reflected in these constraints in the form that each user may request certain requirements.

Maximizing the sum-rate requires a covariance optimization and an efficient user selection. This is a difficult task, specially if the number of antennas is huge or there are many users in the scenario. Moreover, it requires full CSI, which means the BS should know $\sum_k r_k \times M\times t$ parameters.
Even with reduced complexity algorithms, the implementation aspects related to the encoding/decoding of signals and the signalling required to exchange the CSI make the optimal design unfeasible to implement.

With the aim of reducing system complexity in terms of feedback and algorithms, the next section presents a radio resource allocation framework based on:
\begin{itemize}
\item An application scenario where the user terminals have only one antenna. That is $r_k=1, \forall k$,  $\vc{H}_{k,m}$ is a row-vector and the channel is a MISO BC. It is easy to appreciate that the problem is non-concave with respect to power and beamformers and its solution is still a complex task. Moreover, if there are more users than the available resources, user selection is needed. The optimal solution implies a huge exhaustive search among all possibilities and combinations.

\item Multiuser Opportunistic Beamforming (MOB) strategy, which is further explained in reference \cite{Zor01}. The transmitter uses a fixed precoder to determine a set of $t$ spatial subchannels per subcarrier. Each spatial subchannel or beam is characterized by a spatial signature, which is a $t$-dimensional beamforming vector ${\bf b}_{m,q}$, $m=1,\ldots,M$, $q=1,\ldots,t$. Each set $\{ {\bf b}_{m,1},\ldots,{\bf b}_{m,t}\}$, for $m=1,\ldots,M$, has orthonormal elements which are randomly generated by the transmitter\footnote{In practical applications, as in Long Term Evolution (LTE) systems, the set can be chosen within a table of predefined precoding matrices.}. Each pair $(m,q)$ of subcarriers and beams is assigned to one user or none. The discrete variable or index $u_{m,q}\in\mathbb{K}_0=\{0,1,\ldots,K\}$ is used to indicate which user is scheduled to use beam $q$ on subcarrier $m$ ($0$ indicates that no user is scheduled on this resource). The whole set of these variables is the matrix ${\bf U} \in \mathbb{K}_0^{M\times t}$. The BS transmits on the subcarrier $m$ the signal ${\bf s}_{u,m}$ to user $u=u_{m,q}$ with covariance matrix ${\boldsymbol {\Sigma}}_{u,m}= {\bf b}_{m,q}{\bf b}_{m,q}^H p_{m,q}$, where $p_{m,q}$ is the power assigned to the signal sent on subcarrier $m$ and beam $q$. The whole set of powers is the matrix ${\bf P} \in \left(\mathbb{R}^{+}\right)^{M\times t}$, where $\mathbb{R}^{+}$ is the set of positive real numbers. The transmitter only needs a partial CSI, as shown later, to perform subcarrier, beam and power allocation, leading to feedback simplification.

\item  Ergodic sum-rate optimization, which allows to implement iterative algorithms with iterations along time (for instance one iteration per time slot), as explained in \cite{WonEvaBook} and \cite{Gian08}. Here, this approach is extended to the more complex case of multiantenna transmission. Ergodic framework allows to cope with the uncertainty of the mobile channels. In fact, if $[1,\ldots,N]$ is the time interval of the optimization, under ergodic assumption for random processes in the system and for sufficiently large $N$, given a generic metric, $R[n]$,   the approximation $ (1/N)\sum_n R[n] \approx \EX{R[n]}=\EX{R}=\RR$ holds, where $\RR$ does not depend on time $n$. Hence, within this framework, optimizing $\RR$ means optimizing $R[n]$ over time interval $[1,\ldots,N]$.
Since at each scheduling interval an iteration of the algorithms is executed, the complexity is much lower than that of resource allocation algorithm that have to carry out various iterations per scheduling interval. In this respect, the proposed technique can be a good alternative to standard scheduling procedures in LTE (see  \cite{Surf01}).
Moreover, this optimization framework is also suitable for traffic sources that do not strictly require constant rate. Hence, wireless systems for data-centric transmissions encourage the adoption of this kind of optimization.
\end{itemize}

As introduced, the main goal of the proposed approximated solution for rate balancing in the MISO-OFDM BC is to schedule users on each carrier and beam, and to allocate power efficiently. To achieve that, it departs from the MISO-OFDM BC formulation and it introduces several simplifications in order to decrease the computational complexity and maintain the feasibility. It is important to remark that whilst some publications focuses either on the optimal solution to this problem or on heuristics that lead to a simplified solution, the proposed framework combines the mathematics and the feasible implementation. Next section explains the proposed solution to this problem.

\section{Radio resource allocation problem}

The aim of resource allocation is to dynamically assign radio resources to the different users.
The objective is to determine the optimal values for the matrices ${\bf U}$ and ${\bf P}$, defined in the previous section, over the optimizing time interval. The problem (\ref{eq:ratebal}) can be detailed as follows
\begin{equation}
\begin{split}
\label{eq:primprob}
& \max_{{\bf U},{\bf P},\Gamma} \Gamma \\
s.t. \;\; & {\RR_k({\bf U},{\bf P})} \geq \phi_k  \Gamma \;\; \forall k \\
& {\PP({\bf U},{\bf P})} \leq \bar{\mathcal{P}} \\
\end{split}
\end{equation}
with the additional constraint that $p_{m,q}=0$ if $u_{m,q}=0$\footnote{This also means that ${\bf P}$ has an implicit dependence on ${\bf U}$ and vice versa as shown afterwards. However, user scheduling is separated from power allocation to make the optimization problem easier. More details are given in the following sections. }.
 Here,
\begin{equation}
\RR_k({\bf U},{\bf P})= \EX{R_k}=\sum_{m=1}^M \sum_{q=1}^{t}\EX{\delta^{u_{m,q}}_{k}\log_2 \left(1+\gamma_{k,m,q}({\bf P})\right)}
\label{eq:Rk}
\end{equation}
is the achievable average rate for user $k$ under the assumptions outlined aforementioned, where the indicator $\delta^{u_{m,q}}_{k}$ is 1 if the $k$th user allocated on the subcarrier $m$ and beam $q$\footnote{$\delta_k^u$ is the Kronecker's delta: $\delta_k^u=1$ if $u=k$ and $\delta_k^u=0$ otherwise.}. Note that no more than one user can be allocated per beam.
Also,
\begin{equation}
\PP({\bf U},{\bf P})=\sum_{m=1}^M \sum_{q=1}^{t} \EX{p_{m,q}}
\end{equation}
is the total average power assigned to users, which is constrained to the value $\bar{\mathcal{P}}$. Finally,
\begin{equation}
\gamma_{k,m,q}({\bf P})=\frac{p_{m,q}c_{k,m,q}}{1+\sum_{s=1,s\neq q}^{t} p_{m,s}c_{k,m,s} }
\label{eq:gammaini}
\end{equation}
is the Signal-to-Noise-plus-Interference Ratio (SNIR) of user $k$ at frequency $m$ and beam $q$, 
and $c_{k,m,q}=|{\bf h}_{k,m}{\bf b}_{m,q}|^2/\sigma_\omega^2$ is the equivalent channel gain for user $k$ at $m$th carrier through $q$th beam.

This problem belongs to the class of infinite dimensional stochastic programs. The widely known concepts of deterministic optimization like Lagrangian duality, gradient and subgradient search can also be extended to this class of problems.

It is assumed that every $\gamma_{k,m,q}({\bf P})$ are known by the BS. Multi-beam opportunistic design just requires partial CSIT, as in \cite{Zor01}. Orthogonal beams are randomly generated for each subcarrier, where each user estimates the channel and sends a partial CSIT report to the transmitter. Applied to this scheme, this CSIT report should contain the channel gains $c_{k,m,q}$. Nevertheless, it can be further simplified as discussed in further sections.

It is important to underline that in this problem rate and power constraints are referred to average values, which is a relaxation of the instantaneous constraints, leading to a reduction in the complexity of the resulting optimization algorithm.

\section{Dual optimization}

The optimization problem to solve is in general non-convex with respect to ${\bf U}$ and ${\bf P}$. A possible method considered in the literature to handle such problem is to deal with the equivalent convex problem in the dual MAC. After solving the MAC problem, the optimum MAC power can be converted into optimum BC power. This is for instance the approach in \cite{Lat07} or one of the alternatives in \cite{Tej01}. Nevertheless, this approach does not give a solution to user selection. In this work, an algorithm of lower complexity that exploits Lagrangian decomposition \cite{Han05} is proposed, that enables users to adapt their resources locally or in parallel with the aid of limited information exchange,
At the sake of incurring in some error, which is illustrated in the section of results, this approach formulates a dual problem, which is always convex, and therefore common techniques for solving convex problems can be applied.

Before introducing the dual problem, the Lagrangian function is derived as follows
\begin{equation}
\begin{split}
\mathcal{L} & =\Gamma + \sum_{k=1}^{K} \mu_k  \left(\RR_k({\bf U},{\bf P})- \phi_k\Gamma \right) + \lambda  \left(\bar{\mathcal{P}}-\PP({\bf U},{\bf P})\right) \\
&=\Gamma\left( 1-\mub^T\phib\right) +\sum_{k=1}^{K}\left(\mu_k \RR_k({\bf U},{\bf P})-\lambda \PP_k({\bf U},{\bf P})\right)
+ \lambda  \bar{\mathcal{P}}
\end{split}
\label{eq:lagr}
\end{equation}
where the dual variables $\lambda, \mub=[\mu_1,\ldots,\mu_K]^T$ relax the cost function and
\begin{equation}
\PP_k({\bf U},{\bf P})=\sum_{m=1}^M\sum_{q=1}^{t}\EX{\delta_k^{u_{m,q}}p_{m,q}}
\end{equation}
is the power assigned to user $k$. Since the problem is not convex, the KKT conditions are necessary but not sufficient to solve it. For this reason, the dual convex problem is introduced, which gives an upper bound of the primal one. The difference between the dual solution and the primal one is called dual gap, which is zero if both problems are convex.

\subsection{Dual problem}
The dual objective of problem (\ref{eq:primprob}) is defined as
\begin{equation}
g(\lambda,\mub)= \max_{{\bf U},{\bf P},\Gamma} \mathcal{L}
\label{eq:dualobj0}
\end{equation}
which has a feasible solution with respect to $\Gamma$ only when $\mub^T\phib=1 $ (to avoid the case of $\Gamma \rightarrow \infty$). Therefore this factor can be removed and it becomes
\begin{equation}
g(\lambda,\mub)= \max_{{\bf U},{\bf P}} \sum_{k=1}^{K}\left(\mu_k \RR_k({\bf U},{\bf P})-\lambda \PP_k({\bf U},{\bf P})\right)+ \lambda  \bar{\mathcal{P}}.
\label{eq:dualobj}
\end{equation}
This expression decouples the constraints in several variables. The individual rate constraints are decoupled and controlled by the $\mu_k$ variables. On the other side, the power constraint cannot be decoupled among the users and hence they share the same variable $\lambda$. Expression (\ref{eq:dualobj}) also has a minimum with respect to $\mub$ and $\lambda$, which is the upper bound of the primal problem. Thus, the dual problem becomes
\begin{equation}
 \min_{\lambda > 0,\mub \in \mathcal{D}} g(\lambda,\mub),
 \label{eq:dualprob}
\end{equation}
where $\mathcal{D}=\{ \mub \geq\boldsymbol{0}, \; | \mub^T \phib=1\}$.

As the dual problem in (\ref{eq:dualprob}) may not be differentiable, the subgradient search is a good method to solve it. From an initial starting point $(\lambda^0,\mub^0)$, the subgradient search gives a sequence of feasible points according to the following update equations:
\begin{equation}
\begin{split}
\label{eq:multipl}
\lambda^{i+1}&=\left[\lambda^i -\rho^i_\lambda g_\lambda^i\right]^+_{\epsilon}\\
\hat{\mub}^{i+1}&=\left[\mub^i -\rho^i_{\mub} {\bf g}_{\mub}^i\right]^+, \qquad  {\mub}^{i+1}=\hat{\mub}^{i+1}/(\phib^T{\hat{\mub}^{i+1}} )\\
\end{split}
\end{equation}
where $g_\lambda^i$ and ${\bf g}_{\mub}^i=[g_{\mub ,1},\ldots,g_{\mub ,K}]^T $ denote the subgradients of (\ref{eq:dualobj0}) with respect to $\lambda$ and $\mub $, respectively, at the iteration $i$ and can be written as
\begin{equation}
\label{eq:sublambda}
g_\lambda^i= \bar{\mathcal{P}}-\PP({\bf U}^{*i},{\bf P}^{*i}),
\end{equation}
\begin{equation}
\label{eq:submu}
g_{\mub,k}^i= \RR_k({\bf U}^{*i},{\bf P}^{*i})-\phi_k \sum_{s=1}^K \RR_s({\bf U}^{*i},{\bf P}^{*i}), \qquad k=1,\ldots,K,
\end{equation}
where $\mub^{i+1}$ is resized to have ${\mub}^T \phib=1$, $[x]^+=\max(0,x)$, $[x]^+_{\epsilon}=\max(\epsilon,x)$, for an arbitrary small $\epsilon \ :\ 0<\epsilon\ll 1$, and $\rho^i_\lambda$, $\rho^i_{\mub}$ are positive step-size parameters\footnote{The issue of subgradient algorithm convergence is addressed in \cite{WonEvaBook} and \cite{Ber99} which provide conditions on the choice of step-size.}.



In the previous expressions ${\bf U}^{*i}$ and ${\bf P}^{*i}$ indicate the values of functions ${\bf U}$ and ${\bf P}$ for which the Lagrangian function in (\ref{eq:lagr}) is maximized when $\lambda=\lambda^i$, $\mub=\mub^i$. That means, for each iteration in the dual domain, there is an optimal solution in the primal one. Thus, the evaluations of these values require the solution of the problem in (\ref{eq:dualobj}), which is addressed in next subsection. In addition, the evaluation of subgradients requiring the statistical description of the channel is still hard from a practical point of view. For this reason, adaptive algorithms are considered in the following section.

\subsection{Allocation algorithms}

This subsection discusses optimal and suboptimal solutions for the dual objective in (\ref{eq:dualobj0}).
In order to derive the solutions ${\bf U^*,P^*}$ given $\lambda, \mub$, let us consider ${\bf u}_m=[u_{m,1},\ldots,u_{m,t}]^T$ as a row of ${\bf U}$ that contains for $m$th carrier the $t$ user indexes,
and ${\bf p}_m=[p_{m,1},\ldots,p_{m,t}]^T$ as a row of ${\bf P}$ that contains $t$ power values.
The optimal solutions are difficult to be found due the cross dependence between user and power allocation.



However, before discussing the solutions, it is important to note that the formulation of the dual objective can be suitably simplified by considering, as in \cite{WonEvaBook}:\\
i) the exclusivity of subcarrier and slot allocation to users, \\
ii) the separability of average rates and power across subcarriers, \\
iii) the fact that channel gains are identically distributed across subcarriers. \\
\noindent Thus, the dual objective in (\ref{eq:dualobj}) can be evaluated as follows:
\begin{equation}
\begin{split}
g(\lambda,\mub) \; = \; &  M \EX{\max_{{\bf u}_m} \left[ \max_{{\bf p}_m\geq {\bf 0}} \mathcal{M}({\bf u}_m,{\bf p}_m)\right]} +\lambda \bar{\mathcal{P}}
\label{eq:dualobj2}
\end{split}
\end{equation}
with
\begin{equation}
\begin{split}
\mathcal{M}({\bf u}_m,{\bf p}_m)=\sum_{q=1,u_{m,q}\neq 0}^{t}\mu_{u_{m,q}}\log_2 (1+\gamma_{u_{m,q},m,q}({\bf p}_m)) - \lambda {p}_{m,q}.
\label{eq:metric}
\end{split}
\end{equation}
can be solved separately for each single subcarrier. Hence, given $\lambda, \mub$, the solution becomes, for each frequency $m$,
\begin{equation}
{\bf u}_m^*= {\rm arg}\max_{{\bf u}_m} \mathcal{M}^*({\bf u}_m)
\label{eq:u}
\end{equation}
with
\begin{equation}
\mathcal{M}^*({\bf u}_m)= \max_{{\bf p}_m\geq {\bf 0}} \mathcal{M}({\bf u}_m,{\bf p}_m).
\label{eq:M*}
\end{equation}
Here, it is important to remind that the argument of maximization has the implicit constraint $p_{m,q}=0$ if $u_{m,q}=0$, and that the optimal solution ${\bf p}^*_m$ is the argument that finally leads to $\mathcal{M}^*({\bf u}^*_m)$.

Equations (\ref{eq:u}) and (\ref{eq:M*}) highlight the interdependency of two problems: the the space and frequency allocation to user in (\ref{eq:u}) and the power allocation problem in (\ref{eq:M*}). As commented at the beginning of the section, power allocation does not appear as a convex optimization. Following the discussion in \cite{Chi07} a convex formulation can be obtained for problem (\ref{eq:M*}) by approximating the expressions of rates to those in high SNIR regime, by using $\log_2(1+x)\approx \log_2(x)$ (see also Appendix B), or successive convex approximation methods \cite{Pap09} can be applied  which converge with some iterations to optimal solution.
Also note that, space-frequency allocation in (\ref{eq:u}) represents a discrete optimization problem which  requires in general an exhaustive search in the space of all possible vectors ${\bf u}_m$ which has cardinality $(K+1)^t$.

By keeping in mind the main task of achieving low-complexity solutions, suboptimal algorithms are proposed and discussed here. The main result that we want to achieve is the separation of space-frequency allocation and power allocation into two subsequent processes or algorithms. From a practical perspective, the space-frequency allocation can be viewed as the scheduling process which assigns users to each space-frequency resource at each time epoch. This is the subject of the next section.

\section{Low-complexity space-frequency and power allocation}

A simple suboptimal solution for power allocation problem in (\ref{eq:M*}) can be easily obtained by differentiating (\ref{eq:metric}) with respect to ${\bf p}_m$ and equating it to zero, under the assumption of constant uniform power $V$ allocated to the interfering beams. The result is the water-filling solution, expressed as
$$
\tilde{p}_{m,q}= \left\{
\begin{array}{ll}
 \left[ \frac{\mu_{u_{m,q}}}{\lambda \ln 2}-\frac{1+\sum_{s=1,s\neq q,u_{m,s}>0}^{t} Vc_{u_{m,s},m,s}}{c_{u_{m,q},m,q}} \right]^+ & {\rm if } \;\; u_{m,q}>0\\
0 & {\rm if } \;\; u_{m,q}=0
\end{array}
\right.
$$
\begin{equation}
= \left\{
\begin{array}{ll}
 \left[ \frac{\mu_{u_{m,q}}}{\lambda \ln 2}-\frac{V}{\gamma_{u_{m,q},m,q}({\bf V}_m)} \right]^+ & {\rm if } \;\; u_{m,q}>0\\
0 & {\rm if } \;\; u_{m,q}=0
\end{array}
\right.
\label{eq:v_tilde}
\end{equation}
where the components of ${\bf V}_m $ are $v_{m,q}=V(1-\delta^{u_{m,q}}_0)$. The power $V$ can be considered as a parameter which estimates the power of interfering beams. In this way, for the space-frequency allocation the simplified solution becomes
\begin{equation}
\tilde{\bf u}_m= {\rm arg}\max_{{\bf u}_m} \mathcal{M}({\bf u}_m,\tilde{\bf p}_m).
\label{eq:u_tilde}
\end{equation}
This solution reduces the computation complexity of power allocation that should be executed inside (\ref{eq:u_tilde}).

Further simplifications have to be introduced to completely decouple space-frequency allocation and power allocation. In fact, the two processes can be made independent through the replacement of $\mathcal{M}({\bf u}_m,\tilde{\bf p}_m)$ in eq. (\ref{eq:u_tilde}) with $\mathcal{M}({\bf u}_m,{\bf V}_m)$, i.e.
 \begin{equation}
\breve{\bf u}_m= {\rm arg}\max_{{\bf u}_m} \mathcal{M}({\bf u}_m,{\bf V}_m)
\label{eq:u_breve}
\end{equation}
and by evaluating (\ref{eq:v_tilde}) or (\ref{eq:M*}), after spatial allocation, using this approximation of $\breve{\bf u}_m$. It is expected that when the number of users $K$ is very large, the selected combination of users is probably spatially orthogonal, which makes the solution of (\ref{eq:v_tilde}) approaching the exact solution. As it is shown hereafter, this decoupling-based approach joined with a suitable greedy search of spatial allocations enables a reduction of feedback requirements.

The main issue for spatial allocation is to reduce the search space, which may be faced by using greedy selection procedures. The simplest among them is the opportunistic selection which also simplifies feedback requirements. In classical opportunistic beamforming, each user selects the best beam, by assuming that all beams are active with a preassigned power, and feeds back the selected beam with its SNIR, while the base station allocates each beam to the best user selected among those competing for that beam. This opportunistic allocation is a suboptimal solution of problem (\ref{eq:u_breve}) and would be attractive due to the reduced feedback requirements, which is discussed in the next subsection.

However, to avoid the losses that allocation schemes for opportunistic beamforming present when the number of users is moderate or low, a reduced complexity spatial allocation algorithm to solve (\ref{eq:u_breve}) is proposed, as done in \cite{Kan08} for a single-carrier system. This algorithm considers the possibility of switching off some beams to maximize $\mathcal{M}({\bf u}_m,{\bf V}_m)$. Here, the approach of \cite{Kan08} is revised by extending it to OFDMA and by discussing its application. 


\section{Space-frequency user allocation: scheduling}

The proposed space-frequency allocation algorithm is described as follows. Let us subdivide, for each subcarrier $m$, the search space $\mathbb{K}_0^{ t}$ into subsets each one characterized by a number $t'_m \in \{1,\ldots,t\}$ of activated beams\footnote{They are the beams with $u_{m,q}\neq 0$} and by an index $j_m=1,\ldots,\binom{t}{t'_m} $, indicating one of the possible dispositions without repetition of $t'_m$  activated beams. It is worth noting that inside each single subset the $t'_m$ terms of the metric $ \mathcal{M}({\bf u}_m,{\bf V}_m)$ are all decoupled. This means that each term with index $q$ in eq. (\ref{eq:metric}) does not depend on $u_{m,s}$, with $s\neq q$, when ${\bf p}_m={\bf V}_m$. This allows to simplify the search of the best $u_{m,q}$ in each subset by using a simple per-beam search. Moreover, inside each single subset the term $\lambda {p}_{m,q}$ becomes $\lambda V$, which is a constant that does not affect the optimization.

As a result, a reduced complexity algorithm to solve problem (\ref{eq:u_breve}) for each subcarrier $m$ can be obtained as follows (the index $m$ from symbols is omitted for the sake of simplicity). Let us denote with ${\cal U}^{(t',j)} \subset \left\{1,\ldots,K\right\}$ and ${\cal S}^{(t',j)} \subseteq \left\{1,\ldots,t\right\}$  the set of allocated users and the  set of activated beams, respectively, for disposition $j$ of $t'$ beams. Let us also denote with $\gamma_{k,q}^{(t',j)}({\bf V})$ the SNIR evaluated for the $j$-th disposition of $t'$ activated beams by using the power vector ${\bf V}=[v_1,\ldots,v_t]^T$, where $v_q=V$ if $q\in{\cal S}^{(t',j)}$, 0 otherwise. By introducing the rate element $r_{k,q}^{(t',j)}=\log_2\left(1+\gamma_{k,q}^{(t',j)}({\bf V})\right) $, the reduced complexity algorithm becomes as Algorithm \ref{alg}.
\begin{algorithm}[!t]
{\small
\begin{algorithmic}[1]
\caption{Beam allocation algorithm}
\label{alg}
\For {$t'=1,\dots,t$}
    \For {$j=1,\dots,\binom{t}{t'}$}
	\State  {${\cal U}^{(t',j)} \longleftarrow \varnothing$, Metric$(t',j) \longleftarrow 0$}
	\ForAll {$q\in {\cal S}^{(t',j)}$}
	      \State {{\bf find }$k^*={\rm arg}\max_{k\notin{\cal U}^{(t',j)}}
                                   \mu_k r_{k,q}^{(t',j)}$}
	      \State {${\cal U}^{(t',j)}\longleftarrow {\cal U}^{(t',j)}\cup\{k^*\}$}
	      \State {Metric$(t',j) \longleftarrow$ Metric$(t',j)+\mu_{k^*} r_{k^*,q}^{(t',j)}$}
	\EndFor
    \EndFor
\EndFor
\State {{\bf find }$(t'^*,j^*)={\rm arg}\max_{(t',j)}$ Metric$(t',j)$}
\State  {${\cal U}^*\longleftarrow {\cal U}^{(t'^*,j^*)} $}
\end{algorithmic}
}
\end{algorithm}
The complexity of the algorithm scales as $t\sum_{t'} \binom{t}{t'}K=t2^tK$, i.e. linearly with $K$, instead of $(K+1)^t$ of the exhaustive search.

\subsection{Low-complexity space-frequency scheduling}

Suboptimal opportunistic selection can be included in this framework if the number $t'$ of activated beams is fixed to the value $\overline{t}$. This value can be fixed after off-line optimization as a function of system configuration, offered traffic and overall channel conditions, or it may be adaptively derived. In this case, each user $k$ can select the  disposition $j_k$ and beam $q_k$ that maximize the SNIR $\gamma_{k,q}^{(\overline{t},j)}({\bf V})$  and thus its rate, given the current CSI at the terminal\footnote{Note that the SNIR depends only on the channel seen by the terminal.}, whereas the BS selects the best disposition of $\overline{t}$ active beams and allocates each beam to the best user selected among those competing for that beam. If the set of users competing for disposition $j$ and beam $q$ is denoted by $D^{}_{q,j}=\{ k \in \{1,\ldots,K\}\colon q_k=q,j_k=j\}$, the algorithm becomes as Algorithm \ref{alg2}.
\begin{algorithm}[!t]
{\small
\begin{algorithmic}[1]
\caption{Beam allocation algorithm with $\overline{t}$ activated beams}
\label{alg2}
\For {$k=1,\dots,K$}
    \State {{\bf find }$(j_k,q_k)={\rm arg}\max_{j,q} \gamma_{k,q}^{(\overline{t},j)}({\bf V})$}
\EndFor
\For {$j=1,\dots,\binom{t}{\overline{t}}$}
	\State  {${\cal U}^{(\overline{t},j)} \longleftarrow \varnothing$, Metric$(j) \longleftarrow 0$}
	\ForAll {$q\in {\cal S}^{(\overline{t},j)}$ }
	      \If {$D^{}_{q,j}-{\cal U}^{(\overline{t},j)}\neq \varnothing$}
	      \State {{\bf find }$k^*={\rm arg}\max_{k\in D^{}_{q,j}-{\cal U}^{(\overline{t},j)}} \mu_k r_{k,q}^{(\overline{t},j)}$}
	      \State {${\cal U}^{(\overline{t},j)}\longleftarrow {\cal U}^{(\overline{t},j)}\cup\{k^*\}$}
	      \State {Metric$(j) \longleftarrow$ Metric$(j)+\mu_{k^*} r_{k^*,q}^{(\overline{t},j)}$}
	      \EndIf
	\EndFor
    \EndFor
\State {{\bf find }$j^*={\rm arg}\max_{j}$ Metric$(j)$}
\State  {${\cal U}^*\longleftarrow {\cal U}^{(\overline{t},j^*)} $}
\end{algorithmic}
}
\end{algorithm}
It is worth noting that the first 3 lines of the algorithm can be executed in a distributed way at user terminals, thus enabling feedback simplification, as shown later.
Classical opportunistic beamforming works with $\overline{t}=t$ and $j_k=j=1$ and does not require the selection of best disposition. Interesting is that, differently from \cite{WonEvaBook}, where multiple beams and interference were not considered, the presented algorithm can be implemented in interference scenarios in order to mitigate interference effects by resorting to spatial and multiuser diversity.

To summarize, a suboptimal resource allocation procedure working within a dual optimization framework is proposed. In this framework, where a suitable algorithm is running to evaluate the dual variables $\lambda,\mub$ to set up rate and power constraints, a scheduling algorithm first assigns space-frequency resources to users, then a simple allocation algorithm finally assigns powers. The performance gap due to suboptimal resource allocation will be illustrated in the numerical results.

\section{Adaptive algorithms}

The evaluation of the statistical averages required to update the subgradients of dual variables $g^i_\lambda$ and ${\bf g}^i_{\mub}$ in the dual framework described previously is a hard task which requires the knowledge of channel statistics. Moreover, the sub-gradient method requires to iterate at each time slot to converge to the solution. Nevertheless, because of the ergodic maximization, many works such as \cite{Won07,KushnerYin} and \cite{Gian08} introduce the stochastic approximation approach to perform these iterations along time ($i$ becomes the time index $n$), and the evaluation of the average power and rate can be done through an adaptive filtering. It reduces the complexity dramatically since only one iteration is done in each time slot and therefore the convergence is achieved.

Let us introduce the dependence of system metrics on the discrete time $n$. In the adaptive implementation, at the $n$th time slot, spatial allocation and power allocation are first performed to derive ${\bf u}^*_m[n],{\bf p}^*_m[n]$ for each subcarrier $m$, given the values of dual variable estimated at time $n$, $\lambda[n]$ and $\mub[n]$. Then the dual variables are updated according to the following equations (see (\ref{eq:multipl}))\footnote{Step-sizes for $\lambda$ and $\mub$ have not the same value to account for the different magnitudes of the dual variables and of their subgradients.}:
\begin{equation}
\begin{split}
\lambda[n+1]&=[\lambda[n] -\rho_{\lambda,n} g_\lambda[n]]^+_{\epsilon}\\
\hat{\mub}[n+1]&=[\mub[n] -\rho_{\mub,n} {\bf g}_{\mub}[n]]^+, \qquad  {\mub}[n+1]=\hat{\mub}[n+1]/(\phib^T{\hat{\mub}[n+1]} )\\
\end{split}
\end{equation}
where the subgradients are obtained by using stochastic approximation. To simplify the notation the following equations are used:
\begin{eqnarray}
&\tilde{\PP}[n]=\sum_{m=1}^M\sum_{q=1}^{t} {p}^*_{m,q}[n]\\
&\tilde{\RR}_k[n]=\sum_{m=1}^M\sum_{q=1}^{t}\delta^{u^*_{m,q}}_{k}\log_2\left(1+\gamma_{u^*_{m,q},m,q}({\bf p}^*_m[n])\right)
\label{eq:RR}
\end{eqnarray}
as the instantaneous values of the estimated total power and user rates, respectively. Therefore, the subgradients (see (\ref{eq:sublambda})) are estimated by filtering the processes $\tilde{\PP}[n]$ and $\tilde{\RR}_k[n]$ as follows
\begin{equation}
\begin{split}
g_\lambda[n+1]&= \alpha_n \left(\bar{\mathcal{P}}-\tilde{\PP}[n] \right) + (1-\alpha_n)g_\lambda[n] \\
g_{\mub,k}[n+1]&= \alpha_n \left( \tilde{\RR}_k[n] -\phi_k \sum_{s=1}^K \tilde{\RR}_s[n] \right) + (1-\alpha_n) g_{\mub,k}[n] 
\end{split}
\label{eq:submu_ad}
\end{equation}
where $\alpha_n$ is a forgetting factor.


The convergence behaviour of the adaptive algorithm depends on the choice of step-size sequences $\rho_{\lambda,n}$, $\rho_{\mub,n}$ and of the factor $\alpha_n$.
If the channel is stationary, a suitable choice with $\rho_{\lambda,n}\longrightarrow 0$, $\rho_{\mub,n}\longrightarrow 0$, $\alpha_n\longrightarrow 0$ allows a convergence with probability 1 (w.p.1) to the solutions $\lambda^*,\mub^*$ of the dual problem. As an example, the choice proposed in \cite{Won07}, $\rho_{\lambda,n}=\rho_{\mub,n}=\beta_n\geq0$, $\sum_{n=0}^\infty\beta_n=\infty$, $\alpha_n\geq0$, $\beta_n/\alpha_n\longrightarrow 0$, $\sum_{n=0}^\infty(\beta_n^2+\alpha_n^2 )<\infty$, leads to convergence w.p.1 according to Theorem 5.1 in \cite{WonEvaBook} if both conditions, $\sum_{n=0}^\infty\beta_n^2\EX{(\bar{\mathcal{P}}-\tilde{\PP}[n])^2}<\infty$ and $\sum_{n=0}^\infty\beta_n^2\EX{(\tilde{\RR}_k[n] -\phi_k \sum_{s=1}^K \tilde{\RR}_s[n])^2}<\infty$, are satisfied (see Appendix \ref{appA} for proof).

However, when the channel is stationary only in the short-term and is not long-term stationary due to shadowing, mobility, activation/deactivation of users, there is the need to use adaptivity for long-term tracking of channel variations. This can be solved using small constant (independent of $n$) step-sizes and the weak convergence concept as in \cite{KushnerYin} has to be applied to characterize tracking behaviour of the algorithm. Additionally, theoretical results in \cite{KushnerYin} provide mild conditions for weak convergence of the algorithm.

\subsection{Feedback complexity}

The feedback requirement is closely related to which opportunistic selection is considered. In random beamforming schemes, after the generation of the set of beams, the base station activates sequentially each beam and users measure the equivalent channel gain $c_{k,m,q}$. When this measurement phase is completed, the users may send to base station all the  $t\times M$ channel gains or CQI's. This is the feedback requirement for each user.

Feedback requirement can be reduced by fixing $t'$ to a value $\overline{t}\leq t$ and by performing opportunistic beam selection at each subcarrier $m$. In this case, each user $k$ which knows $\overline{t}$ selects both the best beam and the best disposition by assuming that all beams are active with a preassigned power and feeds back the selected beam $q_k$, the selected disposition $j_k$ with its SNIR $\gamma_{k,q_k}^{(\overline{t},j_k)}({\bf V})$, while the base station allocates beams to the best users according to Algorithm \ref{alg2}. Hence, feedback requirement is reduced to $3\times M$. In addition, depending on the delay spread of the channel, the number of parameters to feed back can be further reduced by frequency grouping or chunk processing \cite{Ott08}. In the special case where $\overline{t}=t$, such as classical opportunistic beamforming, feedback is reduced to $2\times M$.

The main issue related to the use of opportunistic beam selection with reduced feedback is the fact that SNIR information does not allow the exact evaluation of the actual rate achievable on the channel. In fact, the SNIR $\gamma_{k,q}({\bf V}_m)$ evaluated and fed back by the users is not the one required to evaluate update equation (\ref{eq:RR}), which is $\gamma_{k,q}({\bf p}^*_m[n])$ and depends on the power values selected for each user by power allocation algorithm. Moreover, rate evaluation is also required to determine the coding and modulation format for transmission. This mismatch is a source of additional errors for the adaptive solution and may cause some perturbation to rate balancing  and also transmission errors. On the other hand, all these problems can be overcome by assigning uniform power $V$ to all the active beams.

\subsection{Computational complexity}
\label{sec:complexity}

Low complexity is  one of the goals of the proposed algorithm. Here, it is evaluated and discussed, with reference to the base station. In the first stage, the base station sends pilot signals and users sense each channel. The related complexity in term of operations is $\OO{Mt}$. After the feedback stage, the space-frequency allocation stage requires the computation of all the signal-to-noise ratios $\gamma_{k,q}^{(t',j)}({\bf V})$ and rate elements $r_{k,q}^{(t',j)}$, for each $t'$, $j$, $q$, $k$ and, implicitly for each $m$. This requires in general $\OO{MK2^{t}t}$ evaluations, as pointed out in previous sections for Algorithm \ref{alg}.

When opportunistic beam selection is used by fixing $t'_m=\overline{t}$ the complexity decreases. In the classical opportunistic beamforming with $\overline{t}=t$,  complexity is further reduced to $\OO{MKt}$, as pointed out in previous sections for Algorithm \ref{alg2}. Finally, $\OO{Mt}$ evaluations are required in the power allocation stage to determine, according to equation (\ref{eq:v_tilde}), the power for each subcarrier and beam, and $\OO{MKt}$ operations are required to update all parameters. Table \ref{tab:complexity} shows all steps and their complexity.

In general, complexity has values in the range from $\OO{MKt}$ to $\OO{MK2^{t}t}$, which is linear in $M$ and $K$. This metric could be used to adjust the trade-off between optimization and complexity. Table \ref{tab:complexity} summarizes the computational and feedback complexities for the different schemes.

\begin{table}[ht]
\caption{Complexity table of algorithm.}
 \center{
\begin{tabular}{l|c}\hline
 1. Pooling: BS sends a pilot signal to sense each channel. & $\OO{Mt}$ \\ \hline
 2.a \underline{adaptive ${t}'$} & \\
\textit{Feedback}. Each user sends $t\times M$ parameters: $c_{k,m,q}$ &  \\
Space-frequency allocation & $\OO{MK2^{t}t}$ \\ \hline
 2.b \underline{fixed $t'=\overline{t}$} & \\
\textit{Feedback}. Each user sends $3\times M$ parameters: $q_k,\ j_k$, & \\ $\gamma_{k,q_k}^{(\overline{t},j_k)}({\bf V})$, for each subcarrier &  \\
Space-frequency allocation & $\OO{MK\binom{t}{\overline{t}} t}$ \\ \hline
2.c \underline{$t'=\overline{t}=t$} & \\
\textit{Feedback}. Each user sends $2\times M$ parameters: $q_k$, & \\ $\gamma_{k,q_k}^{(t,1)}({\bf V})$, for each subcarrier &  \\
Space-frequency allocation  & $\OO{MKt}$ \\ \hline
3. Power allocation & $\OO{Mt}$ \\ \hline
4. Updating parameters $\lambda,\ \mub$ & $\OO{MKt}$ \\ \hline
\end{tabular}
}
\label{tab:complexity}
\end{table}

\chapter{Discussing results}

\makebox[16cm][r]{\lq\lq \textit{The profound study of nature is the most fertile source of mathematical discoveries.}\rq\rq}
\makebox[16cm][r]{Jean Baptiste J. Fourier}\\[1cm]

This chapter evaluates the proposed algorithms and compares them with other implementations. First, the convergence and stability are shown for different plots. Second, the algorithms are configured varying their parameters. Finally, the algorithms are compared with other implementations, one of them optimal or near-optimal, used as benchmark. Unless otherwise specified, the results refer to the suboptimal allocation as in (\ref{eq:u_breve}) and (\ref{eq:v_tilde}) with adaptive selection of $t'$ (Algorithm 1).

The results are obtained from a simulation setup that includes multiuser environment, realistic mobile radio channel, MIMO-OFDMA based physical layer, resource allocation and scheduling capabilities. The set of results mainly uses values of LTE typical configurations with $M=72$ useful subcarriers working on a bandwidth of $1.25$ MHz with 15 KHz of subcarrier spacing. Channel model includes time-frequency correlated fast fading. Fast fading on each link of the MIMO broadcast channel is complex Gaussian, independent across antennas and is modeled according to a 3GPP Pedestrian model. This model has a finite number of complex multipath components with fixed delay (delay spread of $2.3 \mu s$) and power gain (average normalized to $1$). Time correlation is obtained according to a Jakes' model with given Doppler bandwidth ($6 Hz$ in the results). At the base station orthogonal beamforming is adopted, where beam vectors change randomly at each frame.  Beamforming weights are obtained as those of a uniform linear array of antenna with half-wavelength spacing. In the simulated system the total average power constraint is fixed to $0dBw$.

\section{Convergence and stability}

\begin{figure}[!ht]
\begin{center}
\centerline{\includegraphics[width=1\linewidth,draft=false]{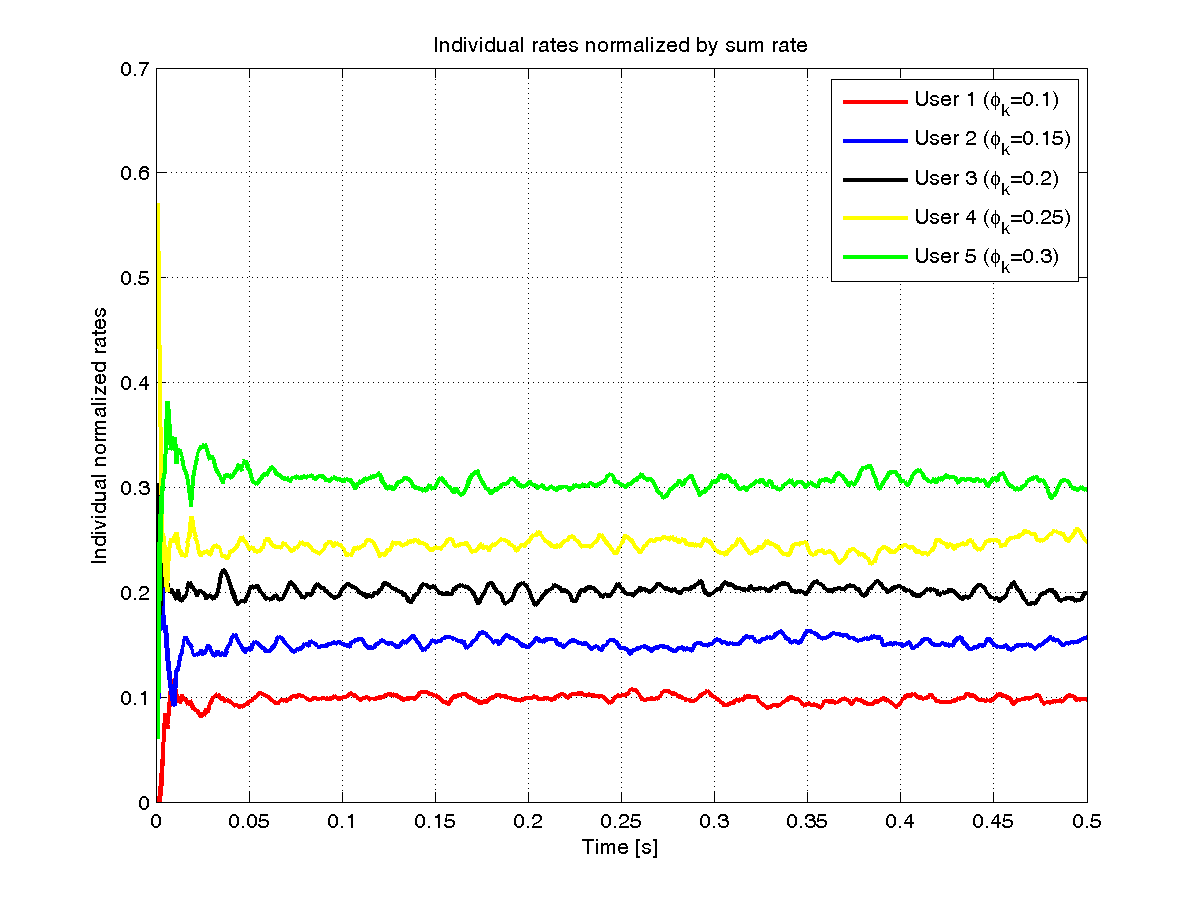}}
\caption{\footnotesize Dynamic behaviour of per-user normalized rates with different values of $\phi_k$. Scenario with users at the same distance and no shadowing and system parameters: $K=5$, $t=4$, $SNR=20 dB$.}
\label{fig_rates_users}
\end{center}
\end{figure}

\begin{figure}[!ht]
\begin{center}
\centerline{\includegraphics[width=1\linewidth,draft=false]{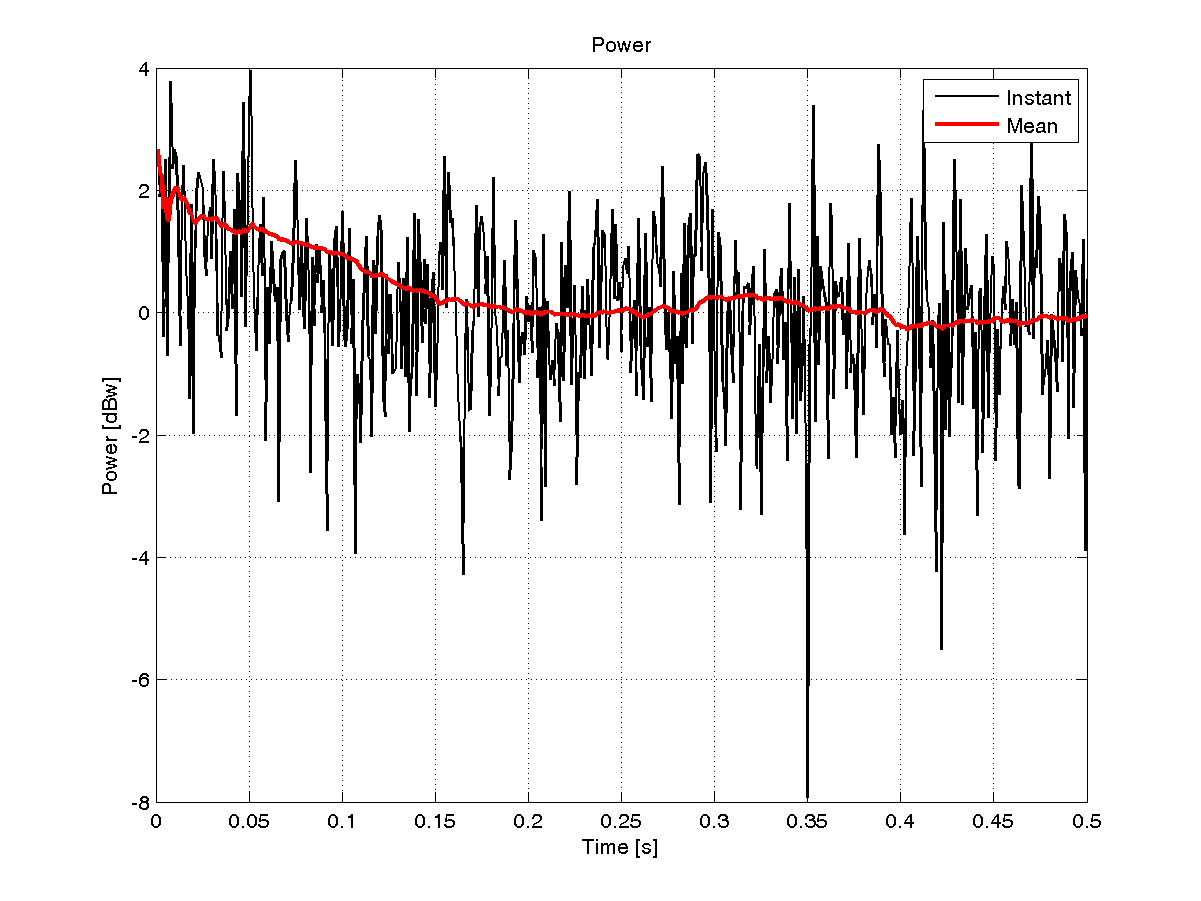}}
\caption{\footnotesize Dynamic behaviour of total power. Scenario and parameters as in Fig.\ref{fig_rates_users}.}
\label{fig_power}
\end{center}
\end{figure}

\begin{figure}[!ht]
\begin{center}
\centerline{\includegraphics[width=1\linewidth,draft=false]{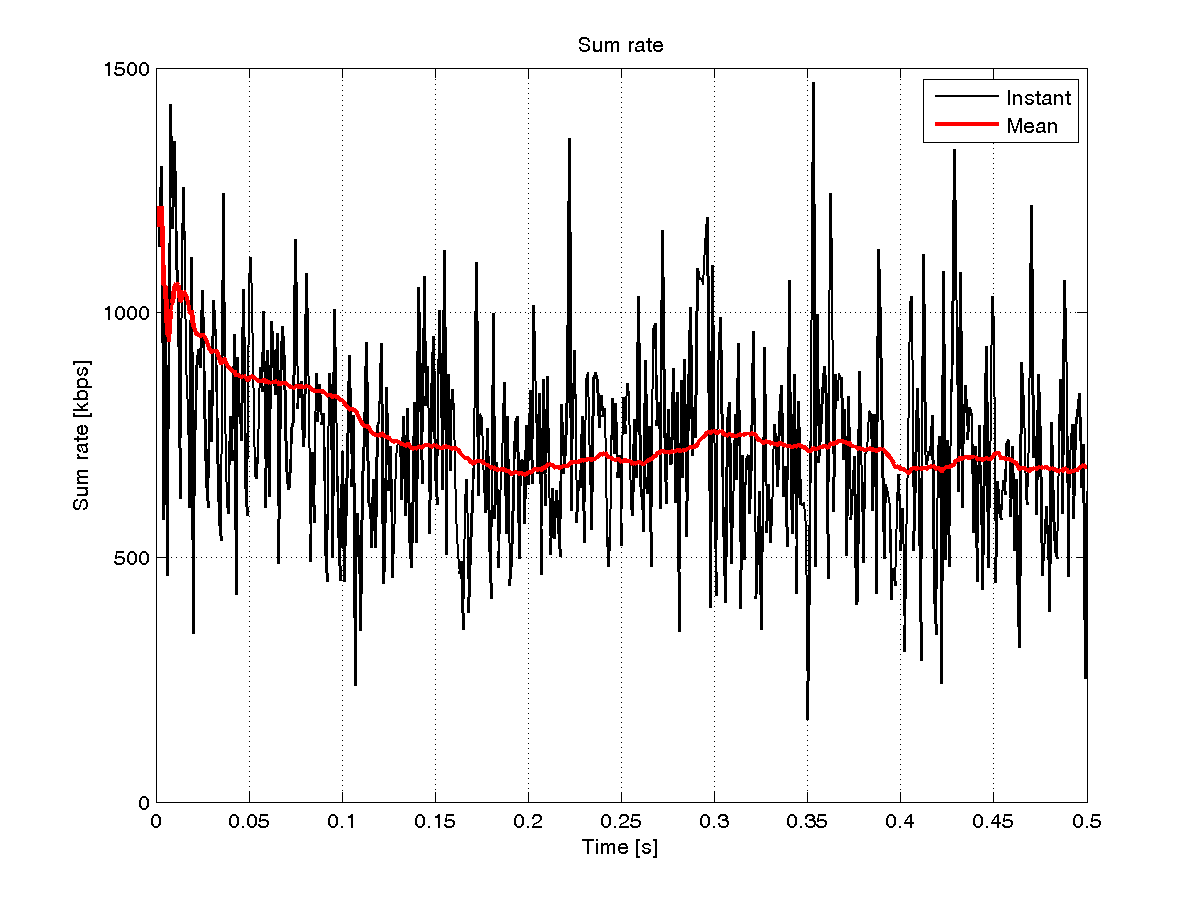}}
\caption{\footnotesize Dynamic behaviour of sum rate. Scenario and parameters as in Fig.\ref{fig_rates_users}.}
\label{fig_sr}
\end{center}
\end{figure}

First, numerical results are obtained in a scenario where users are placed at the same distance from base station and only fast multipath fading is considered for each antenna link. The average (over fading) signal-to-noise ratio per subcarrier at the receivers is denoted as  SNR=$\bar{\mathcal{P}}/(M\sigma_\omega^2)$. Figs.~\ref{fig_rates_users}, \ref{fig_power} and \ref{fig_sr} are obtained for a system with $K=5$ users and  $t=4$ antennas. Users are placed at distance such that SNR=$20$ dB. The figures show the dynamic behaviour of the algorithm and how the average user rates and average total power converge to their final values. The first picture clearly illustrates that the ${\phi}_k$ part of average sum-rate is assigned to user $k$. In the system considered the suboptimal allocation algorithm is implemented.

\section{Evaluation and comparison}

\begin{figure}[!ht]
\begin{center}
\centerline{\includegraphics[width=1\linewidth,draft=false]{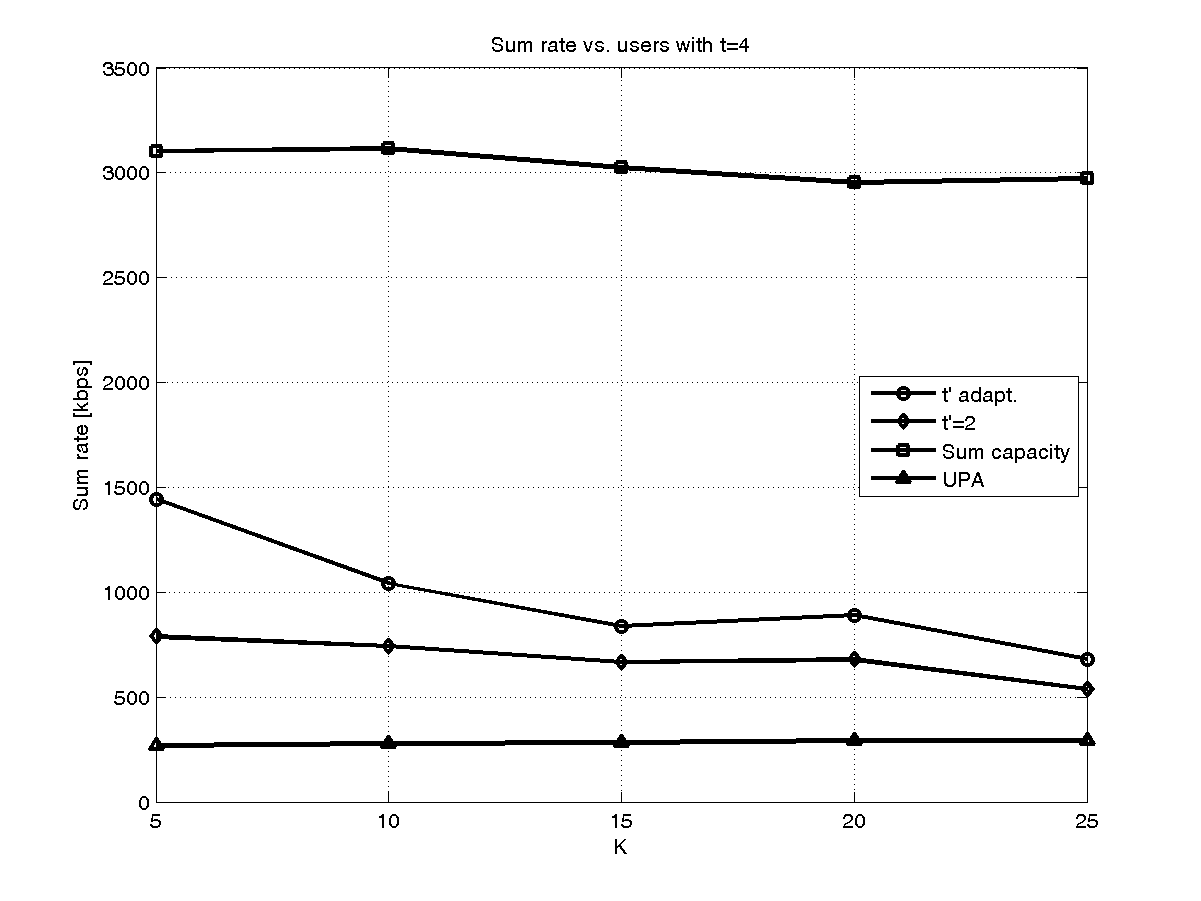}}
\caption{\footnotesize Comparison of different strategies.}
\label{fig:sr_vs_K}
\end{center}
\end{figure}

Fig.~\ref{fig:sr_vs_K} compares the different algorithms described in the work for different numbers $K$ of users. More precisely, the algorithm with optimal power allocation as in (\ref{eq:u}) and (\ref{eq:M*}), the suboptimal algorithm with adaptive selection of $t'$, the suboptimal algorithm with uniform power allocation and constant number $t'=\overline{t}$ of allocated beams (algorithm with reduced feedback) are considered. The figure shows that suboptimal algorithm has a small gap with respect to optimal algorithm and that the schemes with reduced feedback are feasible when the number of allocated beams is small. In this case, the best number of beams, $\overline{t}$, is $2$ if there are at least $4$ antennas. As expected, the schemes with $\overline{t}=t$ have capacity losses, since the available multiuser diversity is not enough to counteract the interference.

\begin{figure}[!ht]
\begin{center}
\centerline{\includegraphics[width=1\linewidth,draft=false]{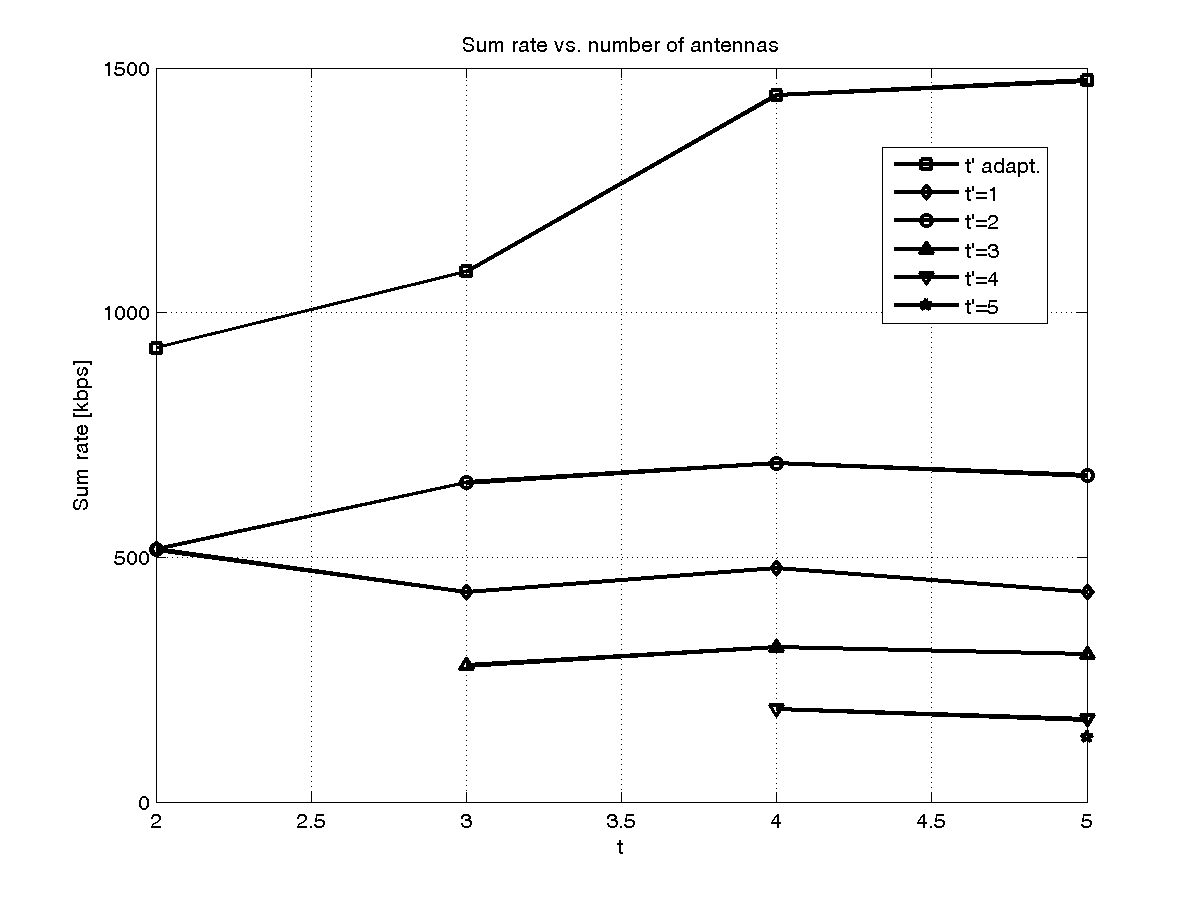}}
\caption{\footnotesize Comparison of different algorithms for different number of antennas.}
\label{fig:sr_vs_t}
\end{center}
\end{figure}

In the same way, fig.~\ref{fig:sr_vs_t} compares the different algorithms described in this work for different numbers of $t$ antennas. As in fig.~\ref{fig:sr_vs_K}, there exists a small gap between adaptive number of active beams and a fixed number of active beams. This picture reveals that the sum rate is increased with the number of antennas with the adaptive $t'$. On the contrary, if a constant $t'$ is chosen, the sum rate does not increase linearly. There is a maximum and it shows that increasing number of antennas may not be suitable. This is explained in terms of interference. Because of fast fading, the optimal number of active beams may vary between accesses since the current beam's weights also vary. For an important number of active beams the interference is also important, it does affect to the sum rate and therefore the system cannot counteract it. Statistically, it is shown that the optimal fixed number of active beams $t'$ is $2$ in the most of cases. 

\begin{figure}[!ht]
\begin{center}
\centerline{\includegraphics[width=1\linewidth,draft=false]{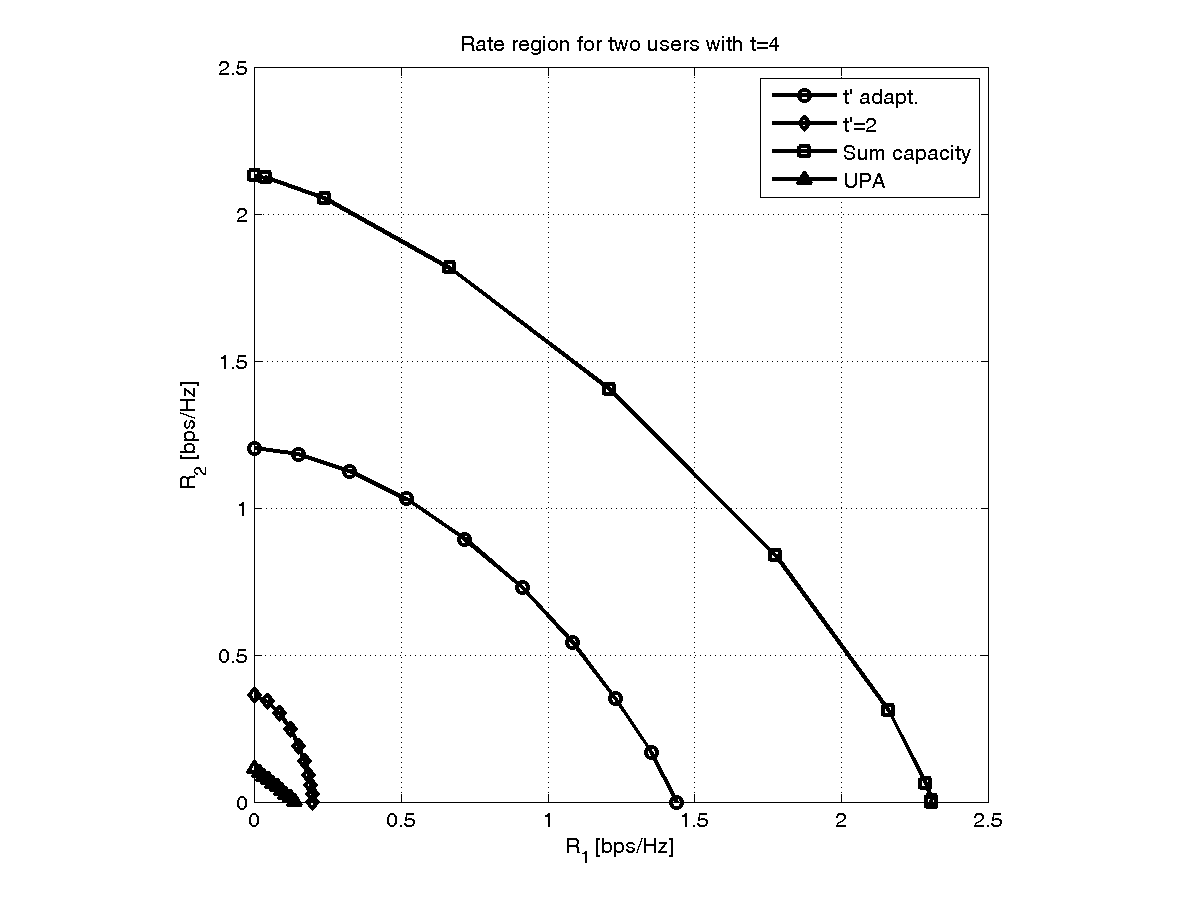}}
\caption{\footnotesize Rate region for the different strategies.}
\label{fig:rate_region}
\end{center}
\end{figure}

Finally, fig.~\ref{fig:rate_region} shows the rate region of two users for the different schemes. Here it is easy to appreciate the gaps between the sum capacity benchmark and the proposed algorithms. While the efficiency is reduced almost to the half, the results of complexity show that the computational complexity is further reduced.

\section{Complexity}

\begin{figure}[!ht]
\begin{center}
\centerline{\includegraphics[width=1\linewidth,draft=false]{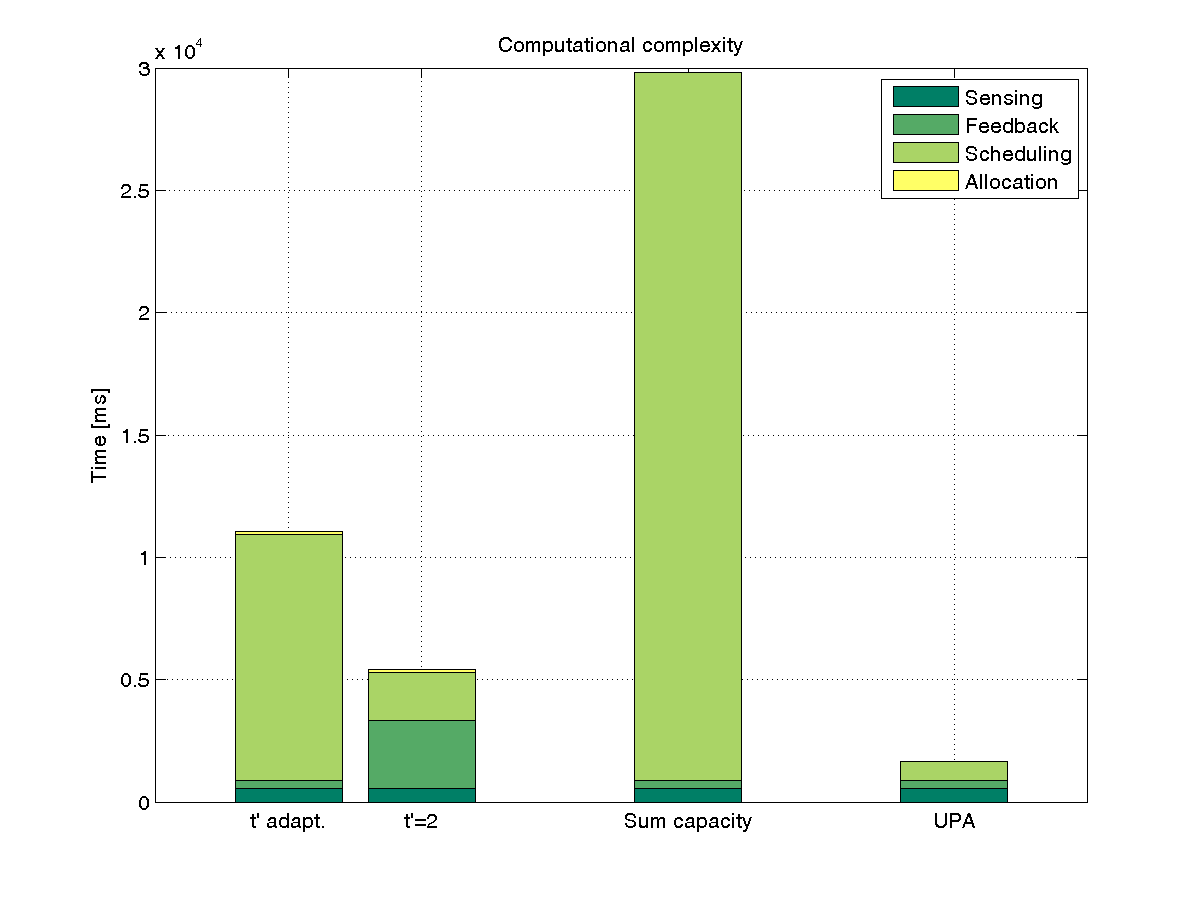}}
\caption{\footnotesize Computational complexity of each stage for different schemes.}
\label{fig:complexity}
\end{center}
\end{figure}

Computational complexity is described in chapter \ref{sec:complexity} and it is an impediment for its implementation. Fig.~\ref{fig:complexity} illustrates the computational complexity of different stages in terms of time consumption for the different schemes. As expected, sum capacity is the most complex, followed by the adaptive $t'$ scheme, even though it is three times lower. The fixed $t'=2$ scheme is half complex and the comparison with figs. \ref{fig:sr_vs_K} and \ref{fig:sr_vs_t} reveals the trade-off between optimization and complexity. 

\begin{figure}[!ht]
\begin{center}
\centerline{\includegraphics[width=1\linewidth,draft=false]{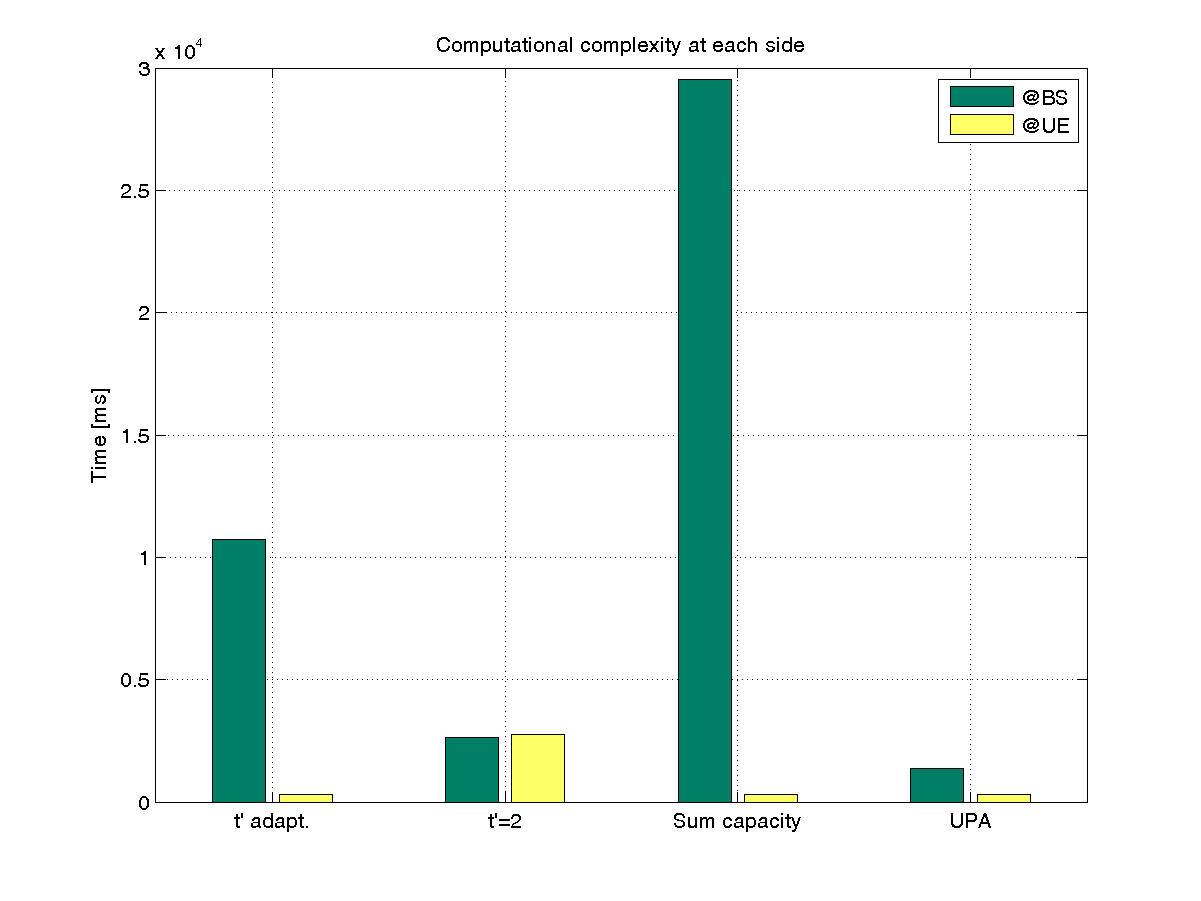}}
\caption{\footnotesize Computational complexity at each side of communication for different schemes.}
\label{fig:complexity_side}
\end{center}
\end{figure}

Furthermore, fig.~\ref{fig:complexity_side} illustrates the overall complexity at BS and UE sides. The most remarkable fact is that the fixed $t'$ reduces the complexity but also distributes the global complexity over all users. Hence, the complexity at base station side is drastically reduced.

It is a remarkable aspect. Both algorithms represent a way for centralizing or distributing complexity among the system. For the first algorithm with dynamic number of beams, the BS processes all information and is able to perform the decision. Hence, the hard task falls into the BS. For the second algorithm, the users performs their decisions and the BS, since there is no cooperative mechanism, performs the final decision. It is evident that some users may choose the same resource and, thus, they will collide on the resource. The BS decides which user grants the resource. The complexity is shared, balanced and distributed between the BS and UEs.

\chapter{Conclusions}

\makebox[16cm][r]{\lq\lq \textit{The five essential entrepreneurial skills for success are concentration,}}
\makebox[16cm][r]{\textit{discrimination, organization, innovation and communication.}\rq\rq}
\makebox[16cm][r]{M. Faraday}\\[1cm]

This work has presented a research based on the common problem in broadcasting scenarios: maximization of the sum rate, with OFDMA and multiantenna schemes. As it has been shown previously, one of the major drawbacks is the dealing with interferences. The other one is the complexity of its implementation. This work aimed to be realistic and feasible, departing from the optimal and theoretical formulation, simplifying it for reducing the complexity and finally, presenting a realistic implementation.

The resulting algorithms comprise all stages involved in the scheduling phases: user selection and resource allocation. Additionally, because of stochastic approximation, the presented strategy can be also implemented as a scheduler of almost typical schemes. Since the algorithms select users, manage resources and stablish priorities, they can be translated into the network as the scheduler of BS. Hence, the advantages are twofold:

\begin{enumerate}
\item Present an efficient mechanism for managing resources.
\item Present the stochastic approximation and ergodic maximization as the scheduling in each time slot without iterative processes.
\end{enumerate}

Finally, the results have reflected the performance of the scheme, compared with other strategies as well as the complexity balanced among the system.

\section{On the trails...}

\subsection{... of fairness}
Fairness is a hot topic in the community. As mentioned, it is studied in several publications, aiming to illustrate an abstract concept as a tangible formula. But which is the tendency? What shall be the next fairness figures? In the introduction some hints have been narrated on that. Currently, in 3G communications fairness is promoted with two classes of users. Each user departs from the A class, which grants users with maximum available rate\footnote{With perfect coverage and current standards, the maximum rate is around ~$7$Mbps.}, and all of them have a certain amount of available data for downloading. Once user reaches its data budget is moved to the second class B. This class limits users with a maximum and lower rate\footnote{Around ~$128$Kbps, depending on the network operator.}. Unlike DSL communications, where the price regulates the speed of the loop, in 3G schemes the price determines the amount of available data for downloading. 

Hence, with this two classes it is easy to appreciate that the fairness is reached when all users achieve what they requested. In other words, the maximum fairness is reached when all classes fulfil their QoS. Probably, the horizon on the fairness figure will be given in terms of achieved objectives and constraints.

\subsection{... of LTE}
As explained, LTE defines many procedures in the different layers of the standard but leaves opened the door to implement mechanisms of selection, resource allocation and feedback interaction. With the feedback parameters defined in the standard may be possible for adapting the present work to LTE. Hence, LTE may be an interesting topic of research and study of its viability.

LTE is a growing topic, not only in the scientific community, but in the industry and it becomes necessary to go further in this direction. 

\chapter*{Appendix A: conditions for convergence}
\label{appA}

Here, it is proved that (i) $\sum_{n=0}^\infty\beta_n^2\EX{(\bar{\mathcal{P}}-\tilde{\PP}[n])^2}<\infty$ and that (ii) \\
$\sum_{n=0}^\infty\beta_n^2$
$\EX{(\tilde{\RR}_k[n] -\phi_k \sum_{s=1}^K \tilde{\RR}_s[n])^2}<\infty$, when step-size $\beta_n$ is such that $\sum_{n=0}^\infty\beta_n^2 <\infty$. These conditions are sufficient to ensure that adaptive algorithm converges to solutions according to theorem 5.1 of \cite{WonEvaBook}.

Let start by stating that
\begin{equation}
 p_{m,q}^*[n]\leq \frac{\max_k\mu_k[n]}{\lambda[n] \ln 2} \leq B <\infty
\end{equation}
where $B$ is a real positive number. If suboptimal allocation algorithm is considered, the inequality is a straightforward derivation of eq. (\ref{eq:v_tilde}). Moreover, this bound is given by a finite number $B$, since $\lambda[n]\geq\epsilon>0$ and ${\mub}^T{\phib}=1$. This bound leads also to the following
\begin{equation}
\log_2\left(1+\gamma_{u^*_{m,q},m,q}\left({\bf p}^*_m[n]\right)\right) \leq \log_2\left(1+c_{u^*_{m,q},m,q}B\right)\leq c_{u^*_{m,q},m,q}\frac{B}{\ln2}.
\end{equation}

The first condition (i) to prove becomes
\begin{equation}
\begin{split}
& \sum_{n=0}^\infty\beta_n^2\EX{\left(\bar{\mathcal{P}}-\tilde{\PP}[n]\right)^2}\\
&\leq\sum_{n=0}^\infty\beta_n^2\bar{\mathcal{P}}^2+\sum_{n=0}^\infty\beta_n^2\EX{\tilde{\PP}[n]^2}\\
&\leq \sum_{n=0}^\infty\beta_n^2\bar{\mathcal{P}}^2+\sum_{n=0}^\infty\beta_n^2 \EX{\left(Mt \max_{m,q}p_{m,q}^*[n]\right)^2}\\
&\leq \sum_{n=0}^\infty\beta_n^2\bar{\mathcal{P}}^2+\sum_{n=0}^\infty\beta_n^2 \left(MtB\right)^2 < \infty.\\
\end{split}
\end{equation}

The second condition (ii) to prove becomes
\begin{equation}
\begin{split}
&\sum_{n=0}^\infty\beta_n^2\EX{\left(\tilde{\RR}_k[n] -\phi_k \sum_{s=1}^K \tilde{\RR}_s[n]\right)^2}\\
&\leq\sum_{n=0}^\infty\beta_n^2\EX{\tilde{\RR}_k[n]^2} +\sum_{n=0}^\infty\beta_n^2\EX{\phi_k^2 \left( \sum_{s=1}^K \tilde{\RR}_s[n]\right)^2}
\end{split}
\end{equation}
where the first term can be expanded as
\begin{equation}
\begin{split}
&\sum_{n=0}^\infty\beta_n^2\EX{\tilde{\RR}_k[n]^2} \\
&\leq \sum_{n=0}^\infty\beta_n^2\EX{\left(\sum_{m=1}^M\sum_{q=1}^{t}\delta^{u^*_{m,q}}_{k}c_{u^*_{m,q},m,q}\frac{B}{\ln2} \right)^2}\\
&\leq \sum_{n=0}^\infty\beta_n^2\left(\frac{B}{\ln2}\right)^2\EX{\left(\sum_{m=1}^M\sum_{q=1}^{t}c_{k,m,q} \right)^2}=\sum_{n=0}^\infty\beta_n^2\left(\frac{B}{\ln2}\right)^2C_k< \infty\\
\end{split}
\end{equation}
and the second term can be rewritten as
\begin{equation}
\begin{split}
&\sum_{n=0}^\infty\beta_n^2\EX{ \left( \sum_{s=1}^K \tilde{\RR}_s[n]\right)^2} \\
&\leq \sum_{n=0}^\infty\beta_n^2\phi_k^2\EX{\left(\sum_{s=1}^K \sum_{m=1}^M\sum_{q=1}^{t}\delta^{u^*_{m,q}}_{s}c_{u^*_{m,q},m,q}\frac{B}{\ln2} \right)^2}\\
&\leq \sum_{n=0}^\infty\beta_n^2\left(\phi_kB/\ln2\right)^2\EX{\left(\sum_{s=1}^K \sum_{m=1}^M\sum_{q=1}^{t}c_{s,m,q}\right)^2}=\sum_{n=0}^\infty\beta_n^2\left(\phi_k\frac{B}{\ln2}\right)^2C< \infty.\\
\end{split}
\end{equation}

Both terms are finite, because the two terms $\EX{\left(\sum_{s=1}^K \sum_{m=1}^M\sum_{q=1}^{t} c_{s,m,q}\right)^2}=C$ and $\EX{\left(\sum_{m=1}^M\sum_{q=1}^{t}c_{k,m,q} \right)^2}=C_k$ are combinations of first-order  and second-order moments of random variables $c_{k,m,q}$, which are finite for the usual fading models of practical interest, as Rayleigh or Rician models.

\addcontentsline{toc}{chapter}{Appendix A}
\bibliographystyle{alpha}
\bibliography{Biblio}

\newcommand{\etalchar}[1]{$^{#1}$}
\begin{thebibliography}{KON{\etalchar{+}}10}

\bibitem[3GP10]{TS36213}
3GPP.
\newblock {\em {TS 36.213: "Evolved Universal Terrestrial Radio Acces (E-UTRA);
  Physical Layer procedures"}}, release 9 edition, 09 2010.

\bibitem[{A. }09]{wsa09}
{A. I. Perez-Neira, P. Henarejos, V. Tralli, and M. A. Lagunas}.
\newblock {A low complexity space-frequency multiuser resource allocation
  algorithm}.
\newblock In {\em IEEE Int. ITG Workshop on Smart Antennas}, 2009.

\bibitem[Ber99]{Ber99}
D.~P. Bertsekas.
\newblock {\em {Nonlinear Programming}}.
\newblock {Athena Scientific}, 2nd edition, 1999.

\bibitem[BN09]{Noss01}
Qing Bai and J.~A. Nossek.
\newblock {A QoS-providing resource allocation scheme in multiuser multicarrier
  systems}.
\newblock In {\em Proc. Second Int. Workshop Cross Layer Design IWCLD '09},
  pages 1--5, 2009.

\bibitem[CJLA07]{Lat07}
M.~Codreanu, M.~Juntti, and M.~Latva-Aho.
\newblock {Low-Complexity Iterative Algorithm for Finding the MIMO-OFDM
  Broadcast Channel Sum Capacity}.
\newblock {\em IEEE Trans. Commun.}, 55(1):48--53, 2007.

\bibitem[Cos83]{Costa1983}
M.~Costa.
\newblock {Writing on dirty paper (Corresp.)}.
\newblock {\em IEEE Trans. Inf. Theory}, 29(3):439--441, 1983.

\bibitem[CTP{\etalchar{+}}07]{Chi07}
M.~Chiang, Chee~Wei Tan, D.~P. Palomar, D.~O'Neill, and D.~Julian.
\newblock {Power Control By Geometric Programming}.
\newblock {\em IEEE Trans. Wireless Commun.}, 6(7):2640--2651, 2007.

\bibitem[CZH05]{Han05}
M.~Chiang, S.~Zhang, and P.~Hande.
\newblock {Distributed rate allocation for inelastic flows: optimization
  frameworks, optimality conditions, and optimal algorithms}.
\newblock In {\em Proc. IEEE 24th Annual Joint Conf. of the IEEE Computer and
  Communications Societies INFOCOM 2005}, volume~4, pages 2679--2690, 2005.

\bibitem[{H. }03]{KushnerYin}
{H. J. Kushner, and G. G. Yin}.
\newblock {\em {Stochastic approximation and recursive algorithms and
  applications}}.
\newblock Springer, 2003.

\bibitem[{H. }06]{Cai01}
{H. Bolcksei, D. Gesbert, C. B. Papadias, and A. van der Veen}.
\newblock {\em {Space-Time Wireless Systems: From Array Processing to MIMO
  Communications}}.
\newblock Cambridge Press, June 2006.

\bibitem[HL09]{LHo09}
W.~W.~L. Ho and Ying-Chang Liang.
\newblock {Optimal Resource Allocation for Multiuser MIMO-OFDM Systems With
  User Rate Constraints}.
\newblock {\em IEEE Trans. Veh. Technol.}, 58(3):1190--1203, 2009.

\bibitem[{I. }08]{WonEvaBook}
{I. C. Wong, and B. Evans}.
\newblock {\em {Resource Allocation in Multiuser Multicarrier Wireless
  Systems}}.
\newblock Springer, 2008.

\bibitem[INN09]{Ismail2009}
S.~Ismail, Chee~Kyun Ng, and N.~K. Noordin.
\newblock Fairness resource allocation for downlink ofdma systems.
\newblock In {\em Proc. IEEE 9th Malaysia Int Communications (MICC) Conf},
  pages 575--579, 2009.

\bibitem[{J. }09]{Bre09}
{J. Brehmer, and W. Utschick}.
\newblock {Nonconcave Utility Maximization in the MIMO Broadcast Channel}.
\newblock {\em EURASIP Journal on Advances in Signal Processing}, 2009:1--13,
  2009.

\bibitem[JL07]{Joung2007}
Jingon Joung and Y.~H. Lee.
\newblock {Regularized Channel Diagonalization for Multiuser MIMO Downlink
  Using a Modified MMSE Criterion}.
\newblock {\em IEEE Trans. Signal Process.}, 55(4):1573--1579, 2007.

\bibitem[JOSP08]{Ott08}
E.~Jorswieck, B.~Ottersten, A.~Sezgin, and A.~Paulraj.
\newblock {Feedback Reduction in Uplink MIMO OFDM Systems by Chunk
  Optimization}.
\newblock In {\em Proc. IEEE Int. Conf. Communications ICC '08}, pages
  4348--4352, 2008.

\bibitem[KON{\etalchar{+}}10]{Surf01}
I.~Z. Kovacs, L.~G. Ordoez, M.~Navarro, E.~Calvo, and J.~R. Fonollosa.
\newblock {Toward a reconfigurable MIMO downlink air interface and radio
  resource management: the SURFACE concept}.
\newblock {\em IEEE Commun. Mag.}, 48(6):22--29, 2010.

\bibitem[KU05]{Kaya2005}
O.~Kaya and S.~Ulukus.
\newblock {Ergodic sum capacity maximization for CDMA:Optimum resource
  allocation}.
\newblock {\em IEEE Trans Inf Theory}, 51(5):1831--1836, 2005.

\bibitem[LWG06]{Gian06}
Qingwen Liu, Xin Wang, and G.~B. Giannakis.
\newblock {A cross-layer scheduling algorithm with QoS support in wireless
  networks}.
\newblock {\em IEEE Trans. Veh. Technol.}, 55(3):839--847, 2006.

\bibitem[LZJW07]{Wan07}
Guangyi Liu, Jianhua Zhang, Feng Jiang, and Weidong Wang.
\newblock {Joint Spatial and Frequency Proportional Fairness Scheduling for
  MIMO OFDMA Downlink}.
\newblock In {\em Proc. Int. Conf. Wireless Communications, Networking and
  Mobile Computing WiCom 2007}, pages 491--494, 2007.

\bibitem[MKL10]{Moon2010}
Sung-Hyun Moon, Jin-Sung Kim, and Inkyu Lee.
\newblock {Limited Feedback Design for Block Diagonalization MIMO Broadcast
  Channels with User Scheduling}.
\newblock In {\em Proc. IEEE Global Telecommunications Conf. (GLOBECOM 2010)},
  pages 1--5, 2010.

\bibitem[PE09]{Pap09}
J.~Papandriopoulos and J.~S. Evans.
\newblock {SCALE: A Low-Complexity Distributed Protocol for Spectrum Balancing
  in Multiuser DSL Networks}.
\newblock {\em IEEE Trans. Inf. Theory}, 55(8):3711--3724, 2009.

\bibitem[RC11]{Rodrigues2011}
E.~B. Rodrigues and F.~Casadevall.
\newblock Control of the trade-off between resource efficiency and user
  fairness in wireless networks using utility-based adaptive resource
  allocation.
\newblock {\em IEEE Commun Mag}, 49(9):90--98, 2011.

\bibitem[RG08]{Gian01}
A.~Ribeiro and G.~B. Giannakis.
\newblock {Layer separability of wireless networks}.
\newblock In {\em Proc. 42nd Annual Conf. Information Sciences and Systems CISS
  2008}, pages 821--826, 2008.

\bibitem[RKJ84]{Raj84}
William R.~Hawe Rajendra K.~Jain, Dah-Ming W.~Chiu.
\newblock {A Quantitative Measure of Fairness and Discrimination for Resource
  Allocation in Shared Computer System}.
\newblock {\em Eastern Research Lab}, 1:1--36, 1984.

\bibitem[sKmK08]{Kan08}
Tae sung Kang and Hyung myung Kim.
\newblock {Optimal Beam Subset and User Selection for Orthogonal Random
  Beamforming}.
\newblock {\em IEEE Commun. Lett.}, 12(9):636--638, 2008.

\bibitem[SSH04]{Spencer2004}
Q.~H. Spencer, A.~L. Swindlehurst, and M.~Haardt.
\newblock {Zero-forcing methods for downlink spatial multiplexing in multiuser
  MIMO channels}.
\newblock {\em IEEE Trans Signal Process}, 52(2):461--471, 2004.

\bibitem[TK06]{Kou06}
I.~Toufik and M.~Kountouris.
\newblock {Power Allocation and Feedback Reduction for MIMO-OFDMA Opportunistic
  Beamforming}.
\newblock In {\em Proc. VTC 2006-Spring Vehicular Technology Conf. IEEE 63rd},
  volume~5, pages 2568--2572, 2006.

\bibitem[TUNB09]{Tej01}
P.~Tejera, W.~Utschick, J.~Nossek, and G.~Bauch.
\newblock {Rate Balancing in Multiuser MIMO OFDM Systems}.
\newblock {\em IEEE Trans. Commun. Technol.*}, 57(5):1370--1380, 2009.

\bibitem[VJG03]{Gol01}
S.~Vishwanath, N.~Jindal, and A.~Goldsmith.
\newblock {Duality, achievable rates, and sum-rate capacity of Gaussian MIMO
  broadcast channels}.
\newblock {\em IEEE Trans. Inf. Theory}, 49(10):2658--2668, 2003.

\bibitem[VT03]{Tse01}
P.~Viswanath and D.~N.~C. Tse.
\newblock {Sum capacity of the vector Gaussian broadcast channel and
  uplink-downlink duality}.
\newblock {\em IEEE Trans. Inf. Theory}, 49(8):1912--1921, 2003.

\bibitem[VTL02]{Tse02}
P.~Viswanath, D.~N.~C. Tse, and R.~Laroia.
\newblock {Opportunistic beamforming using dumb antennas}.
\newblock {\em IEEE Trans. Inf. Theory}, 48(6):1277--1294, 2002.

\bibitem[WE07]{Won07}
I.~C. Wong and B.~L. Evans.
\newblock {Optimal OFDMA Resource Allocation with Linear Complexity to Maximize
  Ergodic Weighted Sum Capacity}.
\newblock In {\em Proc. IEEE Int. Conf. Acoustics, Speech and Signal Processing
  ICASSP 2007}, volume~3, 2007.

\bibitem[WPS{\etalchar{+}}07]{Wei08}
Na~Wei, A.~Pokhariyal, T.~B. Sorensen, T.~E. Kolding, and P.~E. Mogensen.
\newblock {Performance of MIMO with Frequency Domain Packet Scheduling in UTRAN
  LTE Downlink}.
\newblock In {\em Proc. VTC2007-Spring Vehicular Technology Conf. IEEE 65th},
  pages 1177--1181, 2007.

\bibitem[{X. }08]{Gian08}
{X. Wang, and G. B. Giannakis}.
\newblock {Ergodic Capacity and Average Rate-Guaranteed Scheduling for Wireless
  Multiuser OFDM Systems}.
\newblock In {\em Proc. IEEE International Symposium on Information Theory.
  ISIT '08}, volume~1, pages 1691--1695, 2008.

\bibitem[YC04]{Cio02}
Wei Yu and J.~M. Cioffi.
\newblock {Sum capacity of Gaussian vector broadcast channels}.
\newblock {\em IEEE Trans. Inf. Theory}, 50(9):1875--1892, 2004.

\bibitem[YLC04]{WYu01}
Wei Yu, R.~Lui, and R.~Cendrillon.
\newblock {Dual optimization methods for multiuser orthogonal frequency
  division multiplex systems}.
\newblock In {\em Proc. IEEE Global Telecommunications Conf. GLOBECOM '04},
  volume~1, pages 225--229, 2004.

\bibitem[Yu06]{Yu06}
Wei Yu.
\newblock {Uplink-downlink duality via minimax duality}.
\newblock {\em IEEE Trans. Inf. Theory}, 52(2):361--374, 2006.

\bibitem[ZN08]{Zor01}
N.~Zorba and A.~I.~P. Neira.
\newblock {Opportunistic Grassmannian Beamforming for Multiuser and
  Multiantenna Downlink Communications}.
\newblock {\em IEEE Trans. Wireless Commun.}, 7(4):1174--1178, 2008.

\end{thebibliography}
\addcontentsline{toc}{chapter}{Bibliography}

\end{document}